\documentclass[aps,prb,superscriptaddress,twocolumn,showpacs]{revtex4-1}
\usepackage{graphicx,amssymb,amsmath,color}

\newcommand{\Vexc}{V_{\mathrm{exc}}}

\begin{document}

\title{Current noise spectrum of a single particle emitter: theory and experiment}
\date{\today}

\author{F. D. Parmentier, E. Bocquillon, J.-M. Berroir, D. C. Glattli,  B. Pla\c cais and G. F\`eve\\
\emph{\normalsize{Laboratoire Pierre Aigrain, Ecole Normale Sup\'erieure, CNRS (UMR 8551), Universit\'{e} P. et M. Curie, Universit\'{e} D. Diderot,}
\normalsize{24 rue Lhomond, 75231 Paris Cedex 05, France}}\\
and\\
M. Albert, C. Flindt and M. B\"uttiker\\
\emph{\normalsize{D\'epartement de Physique Th\'eorique, Universit\'e de Gen\`{e}ve, CH-1211, Gen\`{e}ve, Switzerland}}\\
}

\begin{abstract}
The controlled and accurate emission of coherent electronic wave packets is of prime importance for future applications of nano-scale electronics.
Here we present a theoretical and experimental analysis of the finite-frequency noise spectrum of a periodically driven single electron emitter. The electron source consists of a mesoscopic capacitor that emits single electrons and holes into a chiral edge state of a quantum Hall sample. We compare experimental results with two complementary theoretical descriptions: On one hand, the Floquet scattering theory which leads to accurate numerical results for the noise spectrum under all relevant operating conditions. On the other hand, a semi-classical model which enables us to develop an analytic description of the main sources of noise when the emitter is operated under optimal conditions. We find excellent agreement between experiment and theory. Importantly, the noise spectrum provides us with an accurate description and characterization of the mesoscopic capacitor when operated as a periodic single electron emitter.
\end{abstract}


%73.23.-b  Electronic transport in mesoscopic systems
%73.63.-b   Electronic transport in nanoscale materials and structures
%72.70.+m Noise processes and phenomena

\maketitle

\section{Introduction}

\begin{figure}
\centering
\includegraphics[width=0.4\textwidth]{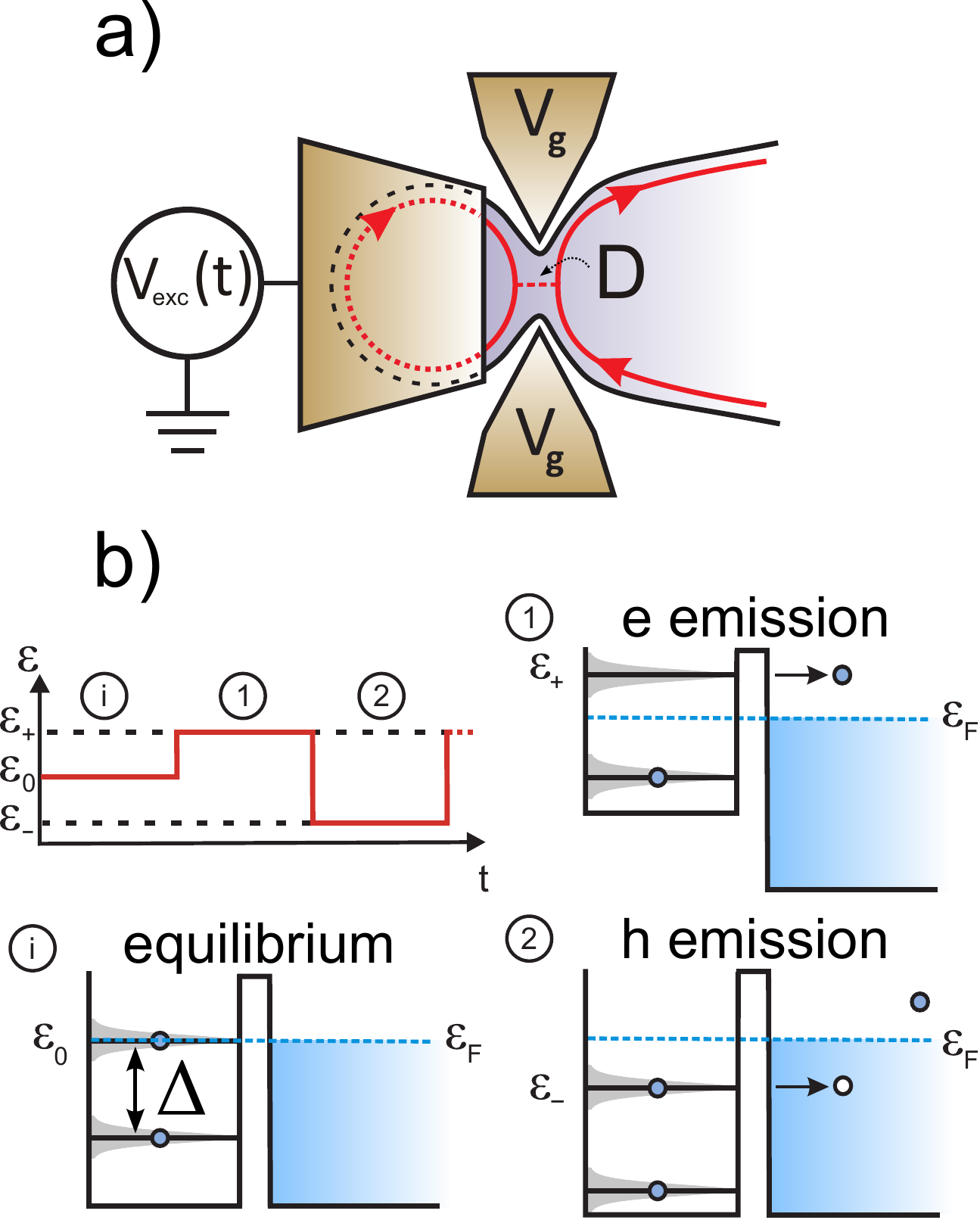}
\caption{(color online). The mesoscopic capacitor. {\bf a)}~Schematic representation of the mesoscopic capacitor. A quantum
point contact (with gate voltage $V_\mathrm{g}$ controlling the
transparency $D$) couples the edge channels (red arrows) in the
capacitor and in the electron gas. Electron/hole emission is
triggered by the excitation drive $V_{exc}(t)$ applied to the
capacitor top gate. {\bf b)} Principle of single charge emission
(in the non-adiabatic regime) using the mesoscopic capacitor. At
\textcircled{\protect\raisebox{-1.2pt}{i}} the capacitor is at
equilibrium and the highest occupied level (HOL) is resonant with
the Fermi level of the electron gas $\epsilon_\mathrm{F}$ (dashed
blue line). The level spacing is denoted as $\Delta$.
\textcircled{\protect\raisebox{-1.2pt}{1}} is electron emission
phase: the HOL is promoted far above the Fermi energy, causing the
capacitor to emit the electron through the quantum point contact.
\textcircled{\protect\raisebox{-1.2pt}{2}} is the hole emission
phase: the emptied level is brought far below the Fermi energy of
the external reservoir and an electron is absorbed from the
electron gas, i.\ e.\ a single hole is emitted. A continuous
repetition of the sequence
\textcircled{\protect\raisebox{-1.2pt}{1}}$\leftrightarrow$\textcircled{\protect\raisebox{-1.2pt}{2}}
results in periodic emission of a single electron followed by a
single hole. This corresponds to the optimal operating condition
of the emitter. } \label{fig-capameso}
\end{figure}

The development of on-demand single electron emitters opens a promising route towards novel nano-scale electronics based on the coherent manipulation of only a single or a few electronic wave packets. The chiral edge channels of the integer quantum Hall effect (IQHE), obtained when a strong perpendicular magnetic field is applied to a two-dimensional electron gas, constitute an ideal experimental setup to test these new concepts in the design of quantum electronic circuits. It is now possible to fabricate micron-sized electrical networks in which the propagation of electrons is truly one-dimensional, ballistic, and quantum coherent, thus mimicking the propagation of photons in optical fibers. Moreover, by depositing metallic split gates on top of the electron gas, quantum point contacts acting as tunable electron beam splitters can also be implemented. Using continuous particle sources, single electron interferences have already been observed in electronic Mach-Zehnder interferometers \cite{Ji2003, Litvin2007, Roulleau2008}, as well as two-electron interferences in similar systems \cite{Neder,Sam2004, Samuelsson2009}. However, for experiments or applications where the timing of wave packets is important, e.\ g.\ for interference experiments that require particles to arrive simultaneously at the scatterer, continuous sources are not useful as one cannot control the emission time of electrons into the conductor. Continuous sources then need to be replaced by triggered emitters\cite{Feve2007, Blum2007, PekolaNP2008, Giazotto2011,LeichtSST2011,
HermelinNat2011,McNeil2011} that can produce single particle states in a controllable and timed manner.

A prime example of a single electron emitter is the periodically
driven mesoscopic capacitor, consisting of a sub-micron cavity coupled to an edge state. The device was first theoretically proposed by B\"uttiker \emph{et al.} \cite{Buttiker1993} who showed that the relaxation resistance is quantized in units of $h/2e^2$ independently of microscopic details \cite{Nig2006,But2007b,Ringel2008,Mora2010,Hamamoto2010,Splettstoesser2010,Filippone2011,Lee2011}. Experimentally, this prediction was confirmed by Gabelli \emph{et al.} \cite{Gabelli2006}. In a subsequent experiment, F\`{e}ve \emph{et al.} showed that the mesoscopic capacitor, when subject to large periodic gate voltage modulations, can absorb and re-emit single electrons at gigahertz frequencies, generating a quantized AC current \cite{Feve2007}.
Several proposals have been made to coherently manipulate the single
electron states emitted by such a source in order to observe
two-electron interferences \cite{Ol'khovskaya,Juergens2011} or electron
entanglement \cite{Splettstoesser2009}. The use of two-particle
exchange effects has also been suggested as a means to visualize
the single electron states generated by the source in a tomography
protocol \cite{Gre10, TheseGrenier} allowing for a direct characterization of
the interaction between a single electronic excitation and its
environment \cite{Degio2009}. Coherence properties of the single
electron states emitted by the source can also be analyzed by
injecting particles into a Mach-Zehnder interferometer
\cite{Haack2011}. However, in order to facilitate these few-fermion experiments, it is necessary first to accurately
characterize and understand in detail the single electron emission
process. The average current measurements performed so far in
Refs.\ \onlinecite{Feve2007} and \onlinecite{Blum2007} give access
to the average behavior of the source after a large number of
particle emissions, but are not designed to provide
information about the elementary processes involved in a single
emission event. For example, average current measurements cannot
distinguish between the deterministic emission of exactly one
electron followed by one hole during each period of the external
driving and a fluctuating number of emitted particles from cycle
to cycle which still results in the same number of emitted charges
after many periods. The statistical properties of the source may
be characterized by measuring the full counting statistics
\cite{Levitov93,Nazarov2003} of electron emissions, but in this case as well,
the short time behavior of the emitter is not accessible. In
contrast, as recently suggested by some of us, the waiting time
distribution \cite{Albert2011} between individual charge events
would provide information about the short-time physics, but an
actual measurement of the waiting time distribution still remains
an open and experimentally challenging task.

We focus in this work on the short time current-current
correlations (or frequency-dependent noise) as an experimental tool to describe
and characterize the elementary excitations generated by the
mesoscopic capacitor when operated as a periodic single electron
source. In two recent short papers we have separately presented
measurements\cite{Mahe2010} and theory \cite{Albert2010} of the finite-frequency current noise of the
mesoscopic capacitor. In the present work, we expand significantly
on the theoretical calculations of the current noise and compare
the recent measurements of the high-frequency noise
\cite{Mahe2010} to two complementary theoretical descriptions: on
one hand, the full-fledged Floquet scattering theory
\cite{Moskalets2002, Moskalets2007, Moskalets2008} which naturally
applies to periodically driven systems and gives highly accurate
numerical results for the current noise, and on the other hand, a
conceptually simple semi-classical model \cite{Mahe2010,
Albert2010} which shows surprisingly good agreement with the
experiment and additionally provides us with a simple intuitive
picture of the dynamical emission processes. Based on the
excellent agreement between the measurements of the short time
current-current correlations and the two theoretical descriptions
we provide a detailed characterization of the single
electron emitter. In particular, we can identify parameter regimes
for which nearly perfect and deterministic single electron-hole
emission is achieved in every cycle. For details of the experiment
we refer the interested reader to Ref.\ \onlinecite{Mahe2010}.

The paper is now organized as follows: In Sec. \ref{sec:capacitor}
we introduce the mesoscopic capacitor and describe the basic
operating principles that enable periodic emission of
electron-hole pairs into an edge state. Sec. \ref{sec:floquet}
gives an elaborate account of the Floquet scattering theory
applied to the mesoscopic capacitor and we discuss calculations of
the average current and the current noise both for a two- and a
three-terminal configuration. We employ a non-interacting model
which allows us to consider all possible parameter ranges and
operating conditions. Sec. \ref{sec-Idiscussion} compares
theoretical predictions of the Floquet scattering theory with
experimental data of the average current. In Sec.\
\ref{sec:heur_mod} we present a semi-classical model that
describes the mesoscopic capacitor around the operating
conditions, where maximally one charge (electron or hole) is
emitted in each half-cycle. The semi-classical model allows us to
account analytically for two important sources of noise: when
electron and hole emissions become rare, the current fluctuations
are shot-noise like and the noise spectrum is white. In contrast,
close to the optimal operating regime, where exactly one
electron-hole pair is emitted in each cycle, the current
fluctuations are dominated by the randomness of the emission times
within a period, giving rise to phase noise with a Lorentzian-like
noise spectrum. In Sec.\ \ref{sec:noise_see} we exhaustively
compare measurements, Floquet scattering theory, and
the semi-classical model, focusing both on current and noise in
different parameter regimes including the shot noise and phase
noise dominated limits. We discuss the general properties of the
noise of the single electron emitter as well as the deviations
between the Floquet scattering theory and the semi-classical model
when the mesoscopic capacitor is operated away from the optimal
conditions. Finally, in Sec.\ \ref{sec:conc} we present our
concluding remarks.

\section{Mesoscopic capacitor}
\label{sec:capacitor}

The mesoscopic capacitor is depicted in Fig.\ \ref{fig-capameso}.
It consists of a submicron-sized cavity (or quantum dot) coupled
to a two-dimensional electron gas through a quantum point contact
(QPC) whose transparency $D$ is controlled by the gate voltage
$V_\mathrm{g}$. A capacitively coupled metallic top-gate controls
the static offset potential in the dot $V_0$, as well as the
rapidly oscillating component $V_{\mathrm{exc}}(t)$ generated by
radio-frequency excitations. A large perpendicular magnetic field
is applied to the sample, so that electrons propagate along the
one-dimensional chiral edge channels that form due to the IQHE.
The system can in principle be operated at an arbitrary integer
value of the filling factor $\nu$ in the electron gas (typically
$\nu=4$), but in any case only the outer edge channel couples to
the quantum dot. Electrons propagating along the outer edge
channel of the quantum dot experience a discrete energy spectrum
with energy levels that are separated by a constant
level spacing $\Delta$, see Fig.\ \ref{fig-capameso}b. The levels
are broadened by the finite coupling between the quantum dot and
the electron gas, determined by the QPC transmission $D$.
Interaction effects within the quantum dot were not experimentally
observed\cite{Feve2007} and are thus neglected throughout this
work (the absence of Coulomb interactions may be due to the
screening from the large metallic top gate as well as the presence
of the inner edge channels in the quantum dot).

Without the periodic driving $V_{\mathrm{exc}}(t)$ (corresponding
to the equilibrium situation denoted as
\textcircled{\protect\raisebox{-1.2pt}{i}} in Fig.\
\ref{fig-capameso}b), the position of the energy levels with
respect to the Fermi energy is determined by the constant voltage
$V_0$ applied to the top-gate: this top-gate voltage fixes the
position of the highest occupied level $\epsilon_0$ (HOL) with
respect to the Fermi energy $\epsilon_\mathrm{F}$ at equilibrium.
Adding next the pure AC excitation voltage $V_{\mathrm{exc}}(t)$
causes the HOL to be periodically shifted up and down with respect
to its equilibrium position. We consider the situation realized
experimentally \cite{Feve2007,Mahe2010}, where a square shape
excitation is applied, causing sudden shifts of the quantum dot
energy spectrum. The square shape excitation contains a broad
range of Fourier components and is thus a non-adiabatic excitation
with respect to all relevant time and energy scales. If the
peak-to-peak amplitude of the excitation drive $2eV_\mathrm{exc}$
is comparable to the level spacing $\Delta$, the HOL is promoted
to an energy $\epsilon_+$ far above the Fermi level in the first
half-period of the drive (labeled as \textcircled{\raisebox
{-1.2pt}{1}} in Fig.\ \ref{fig-capameso}b), where the electron
occupying the level is then emitted to the electron gas through
the quantum point contact. In the following half-period (labeled
as \textcircled{\raisebox {-1.2pt}{2}}), the emptied level is next
brought to an energy $\epsilon_-$ far below the Fermi energy,
where an electron is absorbed from the electron gas (corresponding
to the emission of a hole as indicated in Fig.\
\ref{fig-capameso}). Repeating the sequence \textcircled{\raisebox
{-1.2pt}{1}}$\leftrightarrow$\textcircled{\raisebox {-1.2pt}{2}}
at a drive frequency of $f_d\approx1$ GHz thus gives rise to
periodic emission of a single electron followed by a single hole.

\begin{figure}
\centering\includegraphics[width=0.43\textwidth]{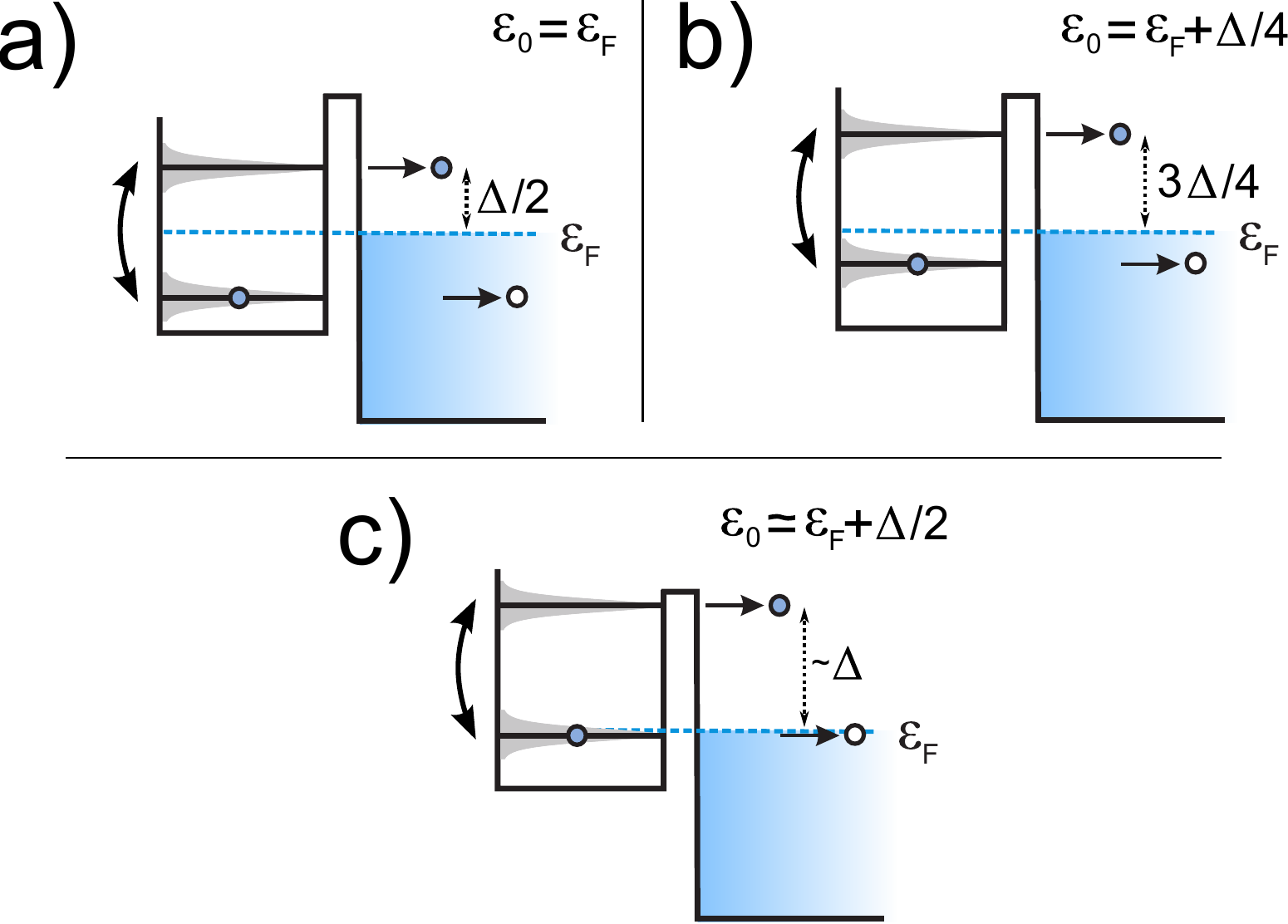}
\caption{(color online). Different operating conditions for the
mesoscopic capacitor. Only the HOL is shown; the curved arrow
represents the AC drive (with amplitude $2e\Vexc=\Delta$) which
periodically and abruptly shifts the HOL above and below the Fermi
level $\epsilon_\mathrm{F}$ (dashed blue line). The dotted arrow
denotes the energy of the emitted electron measured with respect
to the Fermi level. {\bf a)} Optimal operating condition: the energy of the HOL without driving field $\Vexc$
equals the Fermi energy of the electron gas,
$\epsilon_0=\epsilon_\mathrm{F}$. With the driving field applied,
the electron (hole) is then emitted at $\Delta/2$ above (below)
the Fermi level $\epsilon_\mathrm{F}$. {\bf b)} Intermediate case
with $\epsilon_0=\epsilon_\mathrm{F}+\Delta/4$: the electron
(hole) is emitted at $3\Delta/4$ ($\Delta/4$) above (below) the
Fermi level. {\bf c)} Non-optimal operating condition with
$\epsilon_0\approx\epsilon_\mathrm{F}+\Delta/2$: the electron is
then emitted at $\sim\Delta$ above the Fermi level, while the hole
is emitted on resonance with $\epsilon_\mathrm{F}$. In this case,
emissions of spurious electron-hole pairs lead to unwanted
electrical fluctuations.} \label{fig-opcond}
\end{figure}

Obviously, the discussion above depends crucially on the value of
the static potential $V_0$, which fixes the equilibrium position
$\epsilon_0$ of the HOL and thus the positions $\epsilon_+$ and
$\epsilon_-$ during the electron emission (\textcircled{\raisebox
{-1.2pt}{1}}) and hole emission (\textcircled{\raisebox
{-1.2pt}{2}}) phases, respectively, Indeed, as illustrated in
Fig.\ \ref{fig-opcond}, for certain values of $\epsilon_0$, the
HOL may be in resonance with the Fermi level during one of the two
phases. Such a situation is depicted in Fig.\ \ref{fig-opcond}c,
where $\epsilon_0\approx\epsilon_\mathrm{F}+\Delta/2$ and
$2e\Vexc=\Delta$, resulting in
$\epsilon_+\approx\epsilon_\mathrm{F}+\Delta$ and
$\epsilon_-\approx\epsilon_\mathrm{F}$. Thus, during the hole
emission phase, the HOL is resonant with the Fermi level, and
several charges can be absorbed and re-emitted during a single
hole emission phase (note that during the electron emission phase,
the second occupied level is also resonant with the Fermi energy).
As predicted in Ref.\ \onlinecite{Keeling2008}, such emissions of
spurious electron-hole pairs degrade the quality of the single
particle source and lead to unwanted electrical fluctuations. In
this respect, the optimal operating conditions of the emitter are
obtained when the HOL is alternatively brought far above and below
the Fermi level $\epsilon_\mathrm{F}$ during the emission cycle.
Two such cases are shown in Fig.\ \ref{fig-opcond}a and b.
However, even under these favorable conditions, a too large value
of the QPC transmission $D$ may broaden the level so much that it
starts to overlap with the Fermi energy of the lead. The optimal
operating conditions are therefore determined by a subtle
interplay between the static potential $V_0$, the amplitude of the
AC drive $V_\mathrm{exc} $, and the transmission probability $D$
of the QPC.

\section{Floquet scattering theory}
\label{sec:floquet}

We now describe the Floquet scattering matrix theory used to
calculate numerically the average current and the finite-frequency
noise of the periodically driven mesoscopic capacitor. After a
general presentation of the formalism, we apply it to calculate
the current and noise in the experimental situations considered in
this work.

\subsection{Description of the system}

We consider the schematic setup depicted in Fig.\
\ref{fig-floquet}. Electrons in the incoming edge channel can
tunnel onto the quantum dot with the amplitude
$\sqrt{D}=\sqrt{1-r^2}$, perform several round-trips inside the
mesoscopic capacitor, each taking the finite time $\tau_o =l/v_d$,
before finally tunneling back out into the out-going edge state.
In these expressions, the reflection amplitude $r$ has for
convenience been assumed to be real and energy-independent, while
$l$ and $v_d$ are the circumference of the quantum dot and the
drift velocity, respectively. In this setup, the quantum dot acts
as an electronic analog of a Fabry-P\'{e}rot cavity.

We first consider the simple situation, where only a static
potential $V_0$ is applied to the quantum dot. The creation
$\hat{b}^{\dagger}(t)$ and annihilation $\hat{b}(t)$ operators for
an outgoing state at time $t$ are then related to the creation
$\hat{a}^{\dagger}(t')$ and annihilation $\hat{a}(t')$ operators
for an incoming state at time $t'$ through the Fabry-P\'{e}rot
phase acquired after scattering on the quantum dot
\begin{equation}
\begin{split}
    \hat{b}(t) & = \int dt' U(t,t') \hat{a}(t'), \\
        U(t,t') &= r \delta(t- t')  \\
        & - D\sum_{q=1}^{\infty} r^{q-1} \delta(t-t'-q\tau_o) e^{-i e V_0 q\tau_o /\hbar}.
    \end{split}
\end{equation}
Here, $e V_0 \tau_o /\hbar$ is the phase acquired during a single
round-trip inside the quantum dot and $e$ is the electron charge.
In the Fourier domain, the creation $\hat{b}^{\dagger}(\epsilon)$
and annihilation $\hat{b}(\epsilon)$ operators for the outgoing
states are related to the operators of the incoming states
$\hat{a}^{\dagger}(\epsilon)$ and $\hat{a}(\epsilon)$ at energy
$\epsilon=\hbar\omega$ through the stationary scattering matrix
$\mathcal{S}(\epsilon)$:

\begin{figure}
\centering\includegraphics[width=0.27\textwidth]{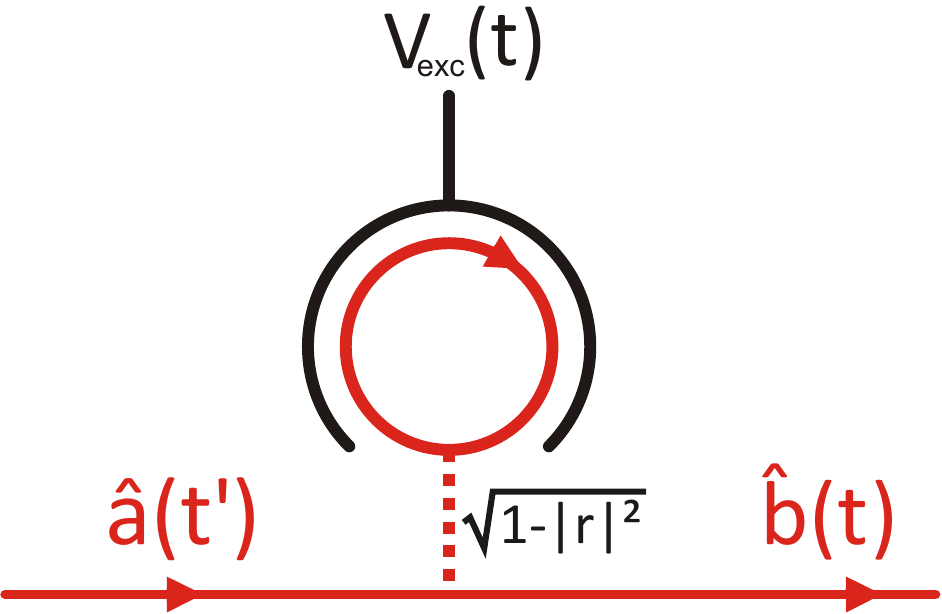}
\caption{(color online). Schematic representation of the
mesoscopic capacitor as a time-dependent scatterer. Electrons in
the in-going edge channel are scattered on the quantum dot (red
loop) subject to the time-dependent potential $V_\mathrm{exc}(t)$.
} \label{fig-floquet}
\end{figure}

\begin{equation}
\begin{split}
\hat{a}(\epsilon) & = \int \frac{dt}{\sqrt{h}} e^{i \epsilon t/\hbar} \hat{a}(t) ,\\
\hat{b}(\epsilon) & = \mathcal{S}(\epsilon) \hat{a}(\epsilon) , \\
\mathcal{S}(\epsilon) &= \frac{r-e^{i \tau_o (\epsilon-\epsilon_0) /\hbar}}{1-r e^{i \tau_o (\epsilon-\epsilon_0) /\hbar}} .
\label{eq:elastic}
\end{split}
\end{equation}
The density of states of the quantum dot\cite{Buttiker1994}
\begin{equation}
\rho(\epsilon) = \frac{1}{2\pi i}
\mathcal{S}^{\ast}(\epsilon)\frac{d\mathcal{S}}{d\epsilon}
\end{equation}
consists of a series of peaks corresponding to the discrete energy
levels of the quantum dot. The typical level spacing $\Delta =
h/\tau_o $ is on the order of a few Kelvins. The levels are
broadened by the coupling to the electron gas with the width of
the peaks given by $D\Delta/ (2 \pi)$. The static potential $\epsilon_0 =
eV_0$ shifts the position of the energy levels measured with
respect to the Fermi energy $\epsilon_\mathrm{F}$ which we thus
freely can set to zero, $\epsilon_\mathrm{F}=0$, throughout the
rest of the paper. The potential shift can be written as a phase
factor $\phi=\epsilon_0 \tau_o  /\hbar$ entering the expression
for the stationary scattering matrix $\mathcal{S}(\epsilon)$. This
allows us to describe the position of the levels in the dot at
equilibrium independently of the level spacing $\Delta$; in
particular, for $\phi=0$ ($\epsilon_0=0$), the highest occupied
level is resonant with the Fermi energy at equilibrium, whereas
for $\phi=\pi$ ($\epsilon_0=\Delta/2$), the Fermi energy lies
midway in between the highest occupied level and the lowest
unoccupied level.

In order to induce a finite AC current in the out-going edge
channel, one must consider a time-dependent modulation of the
quantum dot potential. In addition to the static potential $V_0$,
we therefore consider a periodic modulation $V_\mathrm{exc}(t)$
with no dc component. The period of the modulations is
$T=1/f_d=2\pi/\Omega$, which also defines the drive frequency
$f_d$. The quantum dot can now be viewed as a time-dependent
periodic scatterer, which can be conveniently described using the
Floquet scattering matrix formalism\cite{Moskalets2002}. Between
times $t-q\tau_o$ and $t$, an electron inside the quantum dot
performs $q$ round trips, during which it experiences the
time-dependent potential $V_\mathrm{exc}(t)$. The electron then
acquires an additional phase\cite{Moskalets2008} $\Delta \phi=\frac{e}{\hbar}
\int_{t-q\tau_o}^{t} V_\mathrm{exc}(t')dt'$. Since the drive
$V_\mathrm{exc}(t)$ is periodic, we can express the acquired phase
in terms of the Fourier coefficients $c_n$ entering the Fourier
series

\begin{equation}
e^{-i\frac{e}{\hbar} \int_{0}^{t} V_\mathrm{exc}(t')dt' }=\sum_n c_n e^{-i n \Omega t}.
\end{equation}
The annihilation operators $\hat{b}(\epsilon)$ and $\hat{a}(\epsilon')$ then become related as
\begin{equation}
\begin{split}
 \hat{b}(\epsilon) & = \int d\epsilon' U(\epsilon,\epsilon') \hat{a}(\epsilon'), \\
 U(\epsilon,\epsilon') & = \sum_{n,m}c_n c^{*}_{n+m} \mathcal{S}(\epsilon - n\hbar\Omega) \delta(\epsilon-\epsilon' + m\hbar \Omega)\\
 & = \sum_m U_m(\epsilon) \delta(\epsilon-\epsilon' + m \hbar \Omega).
\end{split}
\end{equation}
The Floquet scattering theory clearly expresses how scattering
occurs through the emission or absorption of a quantized number
$m$ of energy quanta $\hbar\Omega$\cite{Shirley1965}. The
scattering amplitude associated with the transfer of $m$ quanta is
given by the Floquet scattering matrix $U_m(\epsilon)$¨,
\begin{equation}
\begin{split}
\hat{b}(\epsilon) &= \sum_m U_m(\epsilon) \hat{a}(\epsilon_m), \\
U_m(\epsilon)& =  \sum_{n}c_n c^{*}_{n+m} \mathcal{S}(\epsilon_{-n} ), \label{eq-defUm}
\end{split}
\end{equation}
where the notation
\begin{equation}
\epsilon_{\pm m}= \epsilon \pm m \hbar \Omega
\end{equation}
has been introduced for convenience. We note that in the absence
of the excitation drive, only elastic processes can occur and we
recover $U_m(\epsilon)= \mathcal{S}(\epsilon) \delta_{m,0}$ in
agreement with Eq.\ (\ref{eq:elastic}). Finally, from the
unitarity of the time evolution operator $U$, the following
relations can be deduced:
\begin{equation}
\begin{split}
\sum_{n} U_{n+p}^{*}(\epsilon_{-p}) U_{n+p'}(\epsilon_{-p'}) = \delta_{p,p'} \\
\sum_{n} U_{n+p}^{*}(\epsilon_{-n}) U_{n+p'}(\epsilon_{-n}) = \delta_{p,p'} \label{eq-unitarity}
\end{split}
\end{equation}

\subsection{Experimental considerations}

With the Floquet scattering matrices at hand we can now proceed
with calculations of the average current and the finite-frequency
noise. At this point, however, the experimental details of the
measurement setup must be carefully considered. In Fig.\
\ref{fig-23terminal} we show a two and a three-terminal
experimental setup. In the two-terminal geometry, Fig.\
\ref{fig-23terminal}a, the incoming and outgoing edge channels are
connected to the same ohmic contact with a fixed chemical
potential taken as the zero-energy reference
$\mu=\epsilon_\mathrm{F} =0$. The top-gate of the quantum dot is
considered as the second terminal. The two-terminal setup suffices
for measurements of the average current.

However, for measurement of the current noise it is useful to
include an additional ohmic contact as shown in Fig.\
\ref{fig-23terminal}b. The additional ohmic contact is inserted
between the measurement contact and the quantum dot and is
connected to a cold mass. In this geometry, the noise is measured
on a constant impedance, given by the edge channels flowing from
the measurement contact to the grounded contact \cite{GlattliEPJST2009}. The impedance of
the sample seen by the detection circuit is therefore independent
of the parameters of the quantum dot and, in particular, of the
QPC transmission $D$. The environmental noise (noise of the
amplifier or thermal noise emitted by the detection circuit
towards the sample) is reflected on a constant impedance and is
therefore easily subtracted in order to extract the noise emitted
by the source as we discuss in detail below. Only the noise due to
the capacitor is then measured.

\subsubsection{Two-terminal geometry}
\label{subsec:two-term}

\begin{figure}
\centering\includegraphics[width=0.3\textwidth]{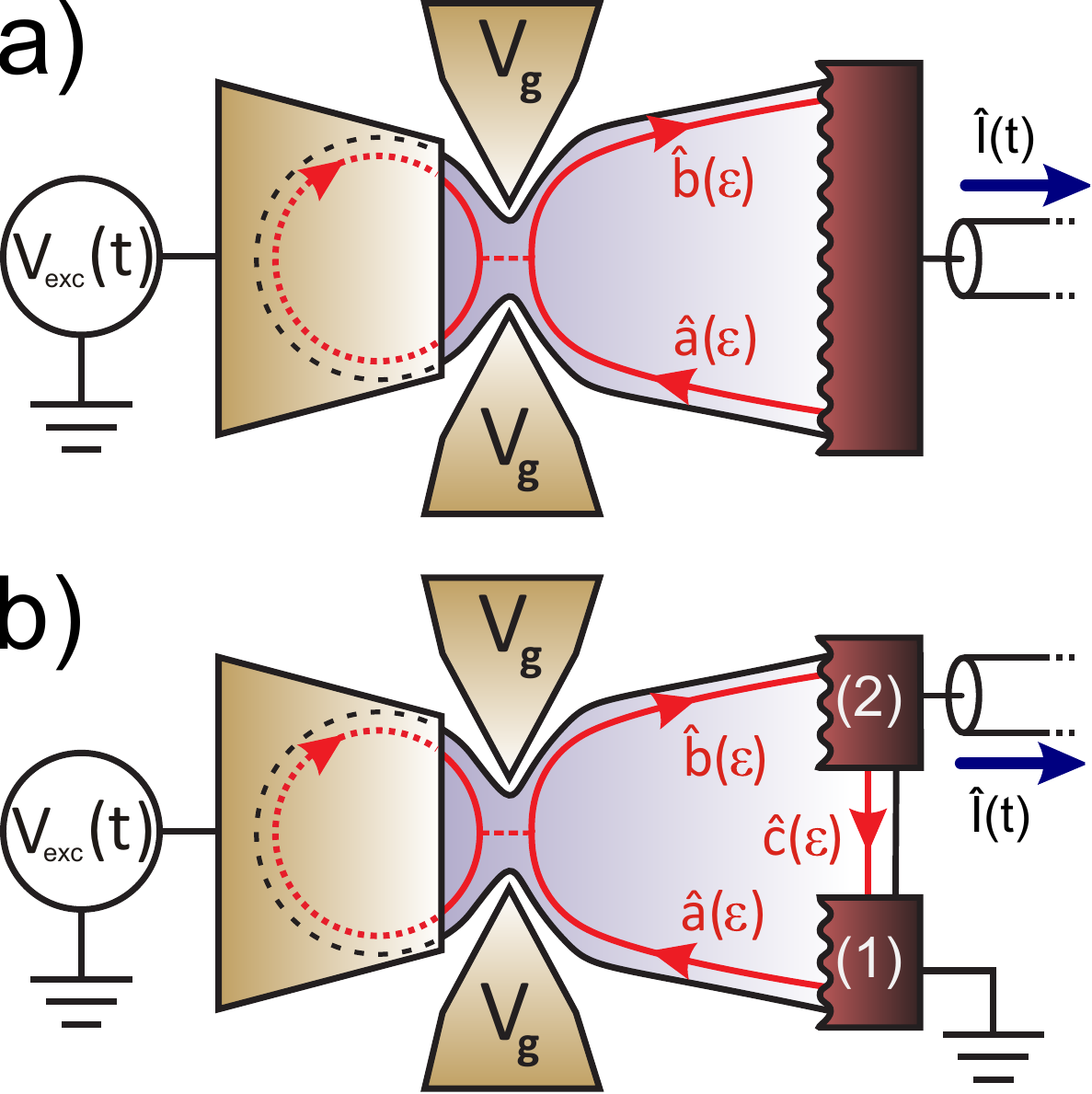}
\caption{(color online). Experimental setups. {\bf a)}
Two-terminal geometry in which the incoming and outgoing edge
channels (red arrows) are both connected to the same ohmic
contact, where the current is measured. {\bf b)} Three-terminal
geometry in which the incoming and outgoing edge states (red
arrows) are connected to ohmic contacts $1$ and $2$, respectively.
The current is measured in contact $2$.} \label{fig-23terminal}
\end{figure}

For the two-terminal geometry in Fig.\ \ref{fig-23terminal}a, the operator for the current emitted by the contact is given as
\begin{equation}
\begin{split}
    \hat{I}(t) & = e \big [ \hat{b}^{\dag}(t)\hat{b}(t) - \hat{a}^{\dag}(t)\hat{a}(t) \big],\\
    & = \frac{e}{h} \int d\epsilon \; d\epsilon' e^{i (\epsilon - \epsilon') t/\hbar} \Big[ \hat{b}^{\dag}(\epsilon)\hat{b}(\epsilon') - \hat{a}^{\dag}(\epsilon)\hat{a}(\epsilon') \Big]\\
    & = \frac{e}{h} \int d\epsilon d\epsilon' \; e^{i (\epsilon - \epsilon')t/\hbar} \\
        & \times \Big[ \sum_{m,m'} U^{*}_{m}(\epsilon)  U_{m'}(\epsilon') a^{\dag} (\epsilon_m) a (\epsilon'_{m'}) - a^{\dag} (\epsilon) a (\epsilon') \Big]. \label{eq-It2term}
\end{split}
\end{equation}
Since the ohmic contact is in thermal equilibrium with temperature $T_\mathrm{el}$ and chemical potential $\mu=0$, we have
\begin{equation}
\langle \hat{a}^{\dag}(\epsilon) \hat{a}(\epsilon') \rangle=  f(\epsilon) \delta(\epsilon-\epsilon'),
\end{equation}
where $f(\epsilon)$ is the Fermi-Dirac distribution. The average current $\langle \hat{I}(t) \rangle$ has the same $T$-periodicity as the driving potential $V_\mathrm{exc}(t)$ and can therefore be written in terms of its Fourier components $I_k$ as
\begin{equation}
\langle \hat{I}(t) \rangle  = \sum_k e^{ik \Omega t} I_k ,\label{eq-avIt2Term}
\end{equation}
where
\begin{equation}
 I_k  =  \frac{e}{h}\!  \sum_\ell\!  \int\! d\epsilon   U^{*}_\ell(\epsilon)U_{\ell+k}(\epsilon_{-k}) \big[f(\epsilon_{\ell})-f(\epsilon) \big]\,. \label{eq-avIk2Term}
\end{equation}
In particular, the first harmonic $I_\Omega=I_{k=1}$ is given by
\begin{equation}
 I_\Omega  =  \frac{e}{h} \sum_\ell  \int d\epsilon   U^{*}_\ell(\epsilon)U_{\ell+1}(\epsilon-\hbar\Omega) \big[f(\epsilon + \ell\hbar \Omega)-f(\epsilon) \big]. \label{eq-avIom2Term}
\end{equation}

We consider next the noise emitted by the source in the two-terminal geometry using the expressions for the current operator $\hat{I}(t)$ in Eq.\ (\ref{eq-It2term}). Since $\hat{I}(t)$ corresponds to a non-stationary current, the current-current correlation function
\begin{equation}
C(t,t')=\langle \delta \hat{I}(t) \delta \hat{I}(t+t')\rangle,
\label{eq:currcurr_corr}
\end{equation}
with $\delta \hat{I}(t)=\hat{I}(t)-\langle \hat{I}(t) \rangle$, depends explicitly both on the absolute time $t$ as well as the time difference $t'$. The current-current correlation function is $T$-periodic in the absolute time $t$, such that it can be expressed in terms of the Fourier components $C_{l}(t')$ as
\begin{equation}
C(t,t') = \sum_{l} e^{il\Omega t} C_{l}(t').
\label{eq:Four_Ser}
\end{equation}
Experimentally, the current noise spectrum is averaged over the absolute time $t$, and only the Fourier component $C_{0}(t')$ is measured. In the frequency domain we have $\mathcal{P}_{0}(\omega) = \int dt' e^{i \omega t'} C_{0}(t')$ with the mean power spectral density for $\omega>0$ defined as
\begin{equation} \label{eq-noisespectrum}
S(\omega)=2\int dt' \overline{\langle \delta I (t) \delta I (t+t')\rangle }^t e^{i\omega t'}=2\mathcal{P}_0(\omega).
\end{equation}
The symbol $\overline{\cdot\cdot\cdot}^t$ denotes averaging over $t$. The noise for the two-terminal geometry $S_{2T}(\omega)$ contains a contribution from the cross-correlations of the current flowing from the contact, $\hat{I}_a(t) = e\hat{a}^{\dag}(t)\hat{a}(t)$, and the current flowing into the contact, $\hat{I}_b(t) = e\hat{b}^{\dag}(t)\hat{b}(t)$ as well as contributions from the autocorrelations of $\hat{I}_a$ and $\hat{I}_b$. The operators $\hat{I}_b$ and $\hat{I}_a$ are related through the Floquet scattering matrix $U_m(\epsilon)$, and after some algebra we arrive at
\begin{widetext}
\begin{equation}
S_{2T}(\omega)=\frac{2e^2}{h}\sum_m \int d\epsilon \left| \delta_{m,0}-\sum_n U^{\ast}_{n}(\epsilon_{-n}) U_{n+m}(\epsilon_{-n}-\hbar\omega) \right|^2 f(\epsilon)\Big(1-f(\epsilon_m+\hbar\omega)\Big) \label{eq-Som2Term}.
\end{equation}
Equations (\ref{eq-avIt2Term}-\ref{eq-avIk2Term}) and (\ref{eq-Som2Term}) are useful for numerical calculations of the average current and the finite-frequency noise, respectively, for the two-terminal geometry.

\subsubsection{Three-terminal geometry}
\label{subsec:three-term}

We next consider the three-terminal geometry depicted in Fig.\
\ref{fig-23terminal}b. In this case, electronic wave packets
incident on the quantum dot have been emitted from contact $1$,
whereas the electronic waves scattering off the quantum dot are
collected in contact $2$. The total current $\hat{I}(t)$ measured
in contact $2$ then reads
\begin{equation}
\begin{split}
\hat{I}(t)  =& e \big [ \hat{b}^{\dag}(t)\hat{b}(t) - \hat{c}^{\dag}(t)\hat{c}(t) \big]\\
=& \frac{e}{h} \! \int \! d\epsilon \, d\epsilon' e^{i (\epsilon - \epsilon') t/\hbar} \Big[ \hat{b}^{\dag}(\epsilon)\hat{b}(\epsilon') - \hat{c}^{\dag}(\epsilon)\hat{c}(\epsilon') \Big],
\end{split}
\end{equation}
where $\hat{c}^{\dag}$ and $\hat{c}$ are the creation and
annihilation operators for edge states between contacts 1 and 2,
see Fig.\ \ref{fig-23terminal}b. Both contacts are in thermal
equilibrium with temperature $T_\mathrm{el}$ and chemical
potentials $\mu_{1,2} = 0$, such that
\begin{equation}
\langle \hat{a}^{\dag}(\epsilon) \hat{a}(\epsilon') \rangle= \langle \hat{c}^{\dag}(\epsilon) \hat{c}(\epsilon') \rangle = f(\epsilon) \delta(\epsilon-\epsilon').
\end{equation}
The average current flowing in contact $2$ can now easily be
calculated and we find that it is exactly equal to the average
current obtained for the two-terminal geometry in Eqs.\
(\ref{eq-avIt2Term}-\ref{eq-avIk2Term}).

The current noise, in contrast, is modified compared to the
two-terminal configuration. In the three-terminal geometry the
operators for the currents flowing to and from contact $2$,
$\hat{I}_b(t)=e\hat{b}^{\dag}(t)\hat{b}(t)$ and
$\hat{I}_c(t)=e\hat{c}^{\dag}(t)\hat{c}(t)$, are independent, such
that their cross-correlation vanishes. We then find
\begin{equation}
S_{3T}(\omega)=\frac{2e^2}{h}\sum_m \int d\epsilon \left[ \delta_{m,0}+\left|\sum_n U^{\ast}_{n}(\epsilon_{-n}) U_{n+m}(\epsilon_{-n}-\hbar\omega) \right|^2 \right] f(\epsilon)\Big( 1-f(\epsilon_m+\hbar\omega) \Big).
\label{eq-Som3Term}
\end{equation}
This expression contains a contribution from the edge channel running from contact $2$ to contact $1$ which is independent of the QPC transmission $D$. In order to remove this noise offset as well as any additional environmental contributions, and thus only to measure the actual noise contribution from the source, we consider the difference $\Delta S_{3T}(\omega)=S_{3T}(\omega,D)-S_{3T}(\omega,D=0)$ between the noise at a given QPC transmission and the noise for $D=0$, where the quantum dot is pinched off. This difference reads
\begin{equation}
\Delta S_{3T}(\omega)=\frac{2e^2}{h}\sum_m \int d\epsilon \left|\sum_n U^{\ast}_{n}(\epsilon_{-n}) U_{n+m}(\epsilon_{-n}-\hbar\omega) \right|^2 f(\epsilon)\Big[f(\epsilon+\hbar\omega)-f(\epsilon_m+\hbar\omega)\Big]\,.
\label{eq-Som3Termdiff}
\end{equation}
\end{widetext}

By construction, $\Delta S_{3T}(\omega)$ vanishes at zero
transmission, $D=0$. Interestingly, it also vanishes at unity
transmission, $D=1$. In this case, we recover the noiseless flow
of charges along a perfectly transmitting
channel\cite{Reznikov1995,Kumar1996,Reydellet2003}. Moreover,
using Eq. (\ref{eq-unitarity}) in (\ref{eq-Som3Termdiff}), we find
$\Delta S_{3T}(\omega=0)=0$. The excess noise generated by the
source has no zero-frequency component and is intrinsically of
finite-frequency nature\cite{Moskalets2008}. Furthermore, it can
be shown that $\Delta S_{3T}(\omega)=\Delta
S_{3T}(-\omega)$\cite{Theseparmentier}, such that the emission and
absorption excess noises are equal. This implies that
$S_{3T}(-\omega)-S_{3T}(+\omega)=2\hbar\omega G_K$, where
$G_K=e^2/h$ is the conductance of the outer edge channel flowing
between the measurement contact and the ground contact. This
result is similar to what was found in Ref.\
\onlinecite{Park2008}. This implies that, as long as only excess
noise $\Delta S_{3T}(\omega)$ is considered, no special care is
required in the ordering of the operators entering the definition
of $C(t,t')$ in Eq.\ (\ref{eq:currcurr_corr}). One would indeed
get the same result for $\Delta S_{3T}(\omega)$ by considering the
inverse ordering of $\hat{I}(t)$ and $\hat{I}(t+t')$ or the
symmetrized time-ordering. From now on, we only consider the
excess noise of the source in the three-terminal geometry $\Delta
S_{3T}(\omega)$, and in order to simplify the notation we define
$S(\omega)\equiv \Delta S_{3T}(\omega)$.

\section{Average current}
\label{sec-Idiscussion}

The expressions for the average AC current $\langle
\hat{I}(t)\rangle$ and $I_\Omega$ given by Eqs.\
(\ref{eq-avIt2Term}-\ref{eq-avIk2Term}) and Eq.\
(\ref{eq-avIom2Term}), respectively, can now be compared to
previous theoretical\cite{Moskalets2007} and
experimental\cite{Gabelli2006,Feve2007,Mahe2008} works. Note
that in the three latter references, comparisons between experimental
data and the time dependent scattering theory were provided. In
these models, however, the AC drive was applied to the ohmic
contacts instead of the top gate of the quantum dot. This
situation differs only by a simple gauge transformation when a
single emitter is considered but would not be applicable to a
system containing several emitters such as those investigated
theoretically in Refs.\
\onlinecite{Ol'khovskaya,Splettstoesser2009}.

\begin{figure}%[!htph]
\centering\includegraphics[width=0.45\textwidth]{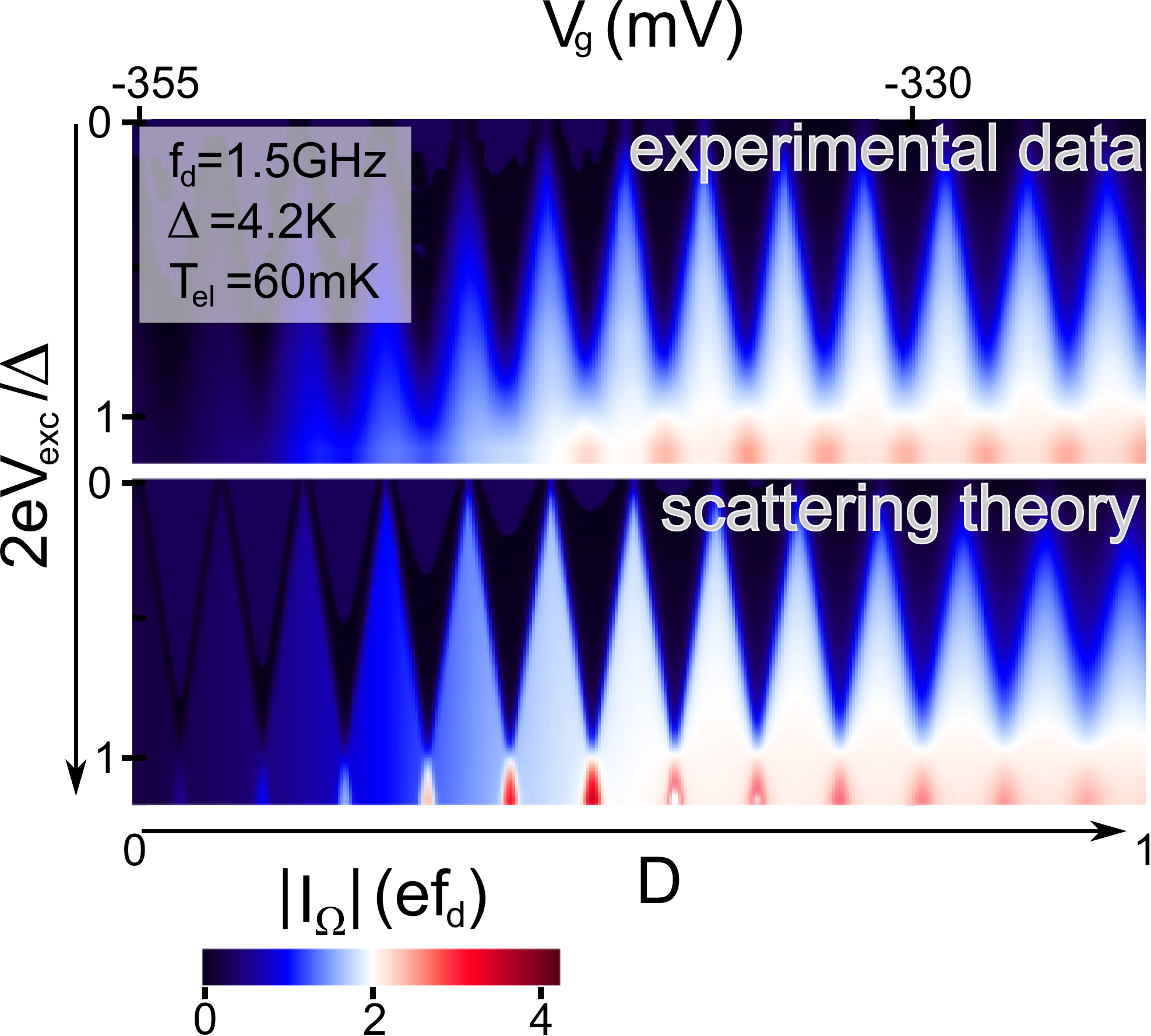}
\caption{(color online). Contour plot of $\left| I_\Omega \right|$
as functions of the square excitation amplitude $2e\Vexc / \Delta$
(vertical scale) and the gate voltage $V_g$ or equivalently the
QPC transmission $D$ (horizontal scale). Top panel: experimental
results for sample $A$ from which we extract the parameters
$\Delta=4.2K$, $f_\mathrm{d}=1.5$ GHz, and $T_\mathrm{el}\approx
60$ mK, see text. Bottom panel: numerical calculations based on
Eq.\ (\ref{eq-avIom2Term}) using the extracted model parameters. }
\label{fig-2DplotACcurrent}
\end{figure}

We now consider actual experimental data: in Fig.\
\ref{fig-2DplotACcurrent}a, we show experimental results for
$|I_\Omega|$ as functions of the square excitation amplitude
$\Vexc$ and the QPC gate voltage $V_\mathrm{g}$ (or
correspondingly the QPC transparency $D$) taken from a sample that
we label as sample $A$. The experiment was carried out in a
dilution refrigerator and the driving frequency was $f_d=1.5$ GHz.
The experimental results display a series of diamond-like
structures centered around $2e\Vexc/\Delta=1$. The spacing of the
diamonds at low gate voltages $V_\mathrm{g}$, where the QPC is
nearly pinched off, allows us to extract the level spacing
$\Delta=4.2$ K of the quantum dot. The electron temperature
$T_\mathrm{el}\approx 60$ mK can moreover be determined from the
smearing of the diamond structures. Finally, the capacitive
coupling between the QPC gates and the quantum dot can be
evaluated from the shift of the position of the levels as $V_g$ is
varied. Using these parameters we evaluate numerically Eq.\
(\ref{eq-avIom2Term}) as shown in Fig.\ \ref{fig-2DplotACcurrent}
(bottom panel). To this end, we use a saddle-point transmission
law for the QPC\cite{Buttiker1990}. The agreement between the
experimental data and numerical calculations is very good, up to
small energy-dependent variations in the QPC transmission which
were not included in the model.

Figure \ref{fig-2DplotACcurrent} allows us to locate operating
regimes for which the mesoscopic capacitor is expected to function
optimally as a controllable single electron source. The white
areas of the diamonds correspond to plateaus on which the current
$\left| I_\Omega \right|=2ef_d$ is given by the product of the
driving frequency and twice the elementary charge $e$ (electrons
and holes each contribute with an elementary charge, resulting in
the factor of 2). In these regions, the device acts \emph{on
average} as a single electron source which emits in each cycle one
single electron (hole) at a well defined energy far above (below)
the Fermi level. The sharpness of the diamonds is then related to
the accuracy of the expected quantization. Note that at low
$\Vexc$ these regions become sharp peaks in $\left| I_\Omega
\right|$ as a function of $V_\mathrm{g}$, corresponding to the HOL
being resonant with the Fermi energy in absence of the driving
($\phi=0$) (in the linear regime $2e\Vexc\ll h f_d, k_B T$, we
also recover theoretically the known expression for the average AC
current \cite{Buttiker1993,Gabelli2006}). In contrast, the values
of $V_\mathrm{g}$, where two diamonds intersect, interpolate at low
$\Vexc$ to zeroes in the conductance corresponding to the Fermi
energy lying midway in between two energy levels ($\phi=\pi$). For
large transmissions, the quantization is gradually lost and the
diamonds fade into a linear dependence of the current on the
driving amplitude. For low transmissions on the other hand, the
typical escape time of an electron on the quantum dot becomes much
longer than the period, and charge emission becomes rare such that
$I_\Omega$ is strongly suppressed.

In the strong driving regime $2e\Vexc\gg h f_d$, we recover
theoretically well-known results for the average AC
current\cite{Feve2007}. In particular, a second-order expansion of
Eq.\ (\ref{eq-avIk2Term}) in the driving frequency, $\hbar\Omega
=hf_d \ll D\Delta$, enables us to express $\langle \hat{I}(t)
\rangle$ as the current response of an $RC$
circuit.\cite{Feve2007} Concretely, we find
\begin{equation}
\langle \hat{I}(t) \rangle=\sum_k   \frac{i\Vexc}{\pi(2k+1)}  \left( -i\Omega C_q +\Omega^2 R_q C^2_q \right) e^{i(2k+1) \Omega t},
\label{eq-Itdev}
\end{equation}
where
\begin{equation}
    C_q  = e^2 \int d\epsilon \rho(\epsilon)  \mathcal{F}(\epsilon,\Vexc)
\end{equation}
and
\begin{equation}
     R_q  = \frac{h}{2e^2} \frac{\int d\epsilon \rho^2(\epsilon) \mathcal{F}(\epsilon,\Vexc)}{\left[ \int d\epsilon \rho(\epsilon)  \mathcal{F}(\epsilon,\Vexc) \right]^2}.  \label{eq-RqDOSNL}
\end{equation}
are the $\Vexc$-dependent capacitance $C_q$ and resistance $R_q$, respectively, and we have defined the function
\begin{equation}
\mathcal{F}(\epsilon,\Vexc)= \frac{f(\epsilon-e\Vexc)-f(\epsilon+e\Vexc)}{2e\Vexc}
\end{equation}
in terms of the Fermi-Dirac distribution $f(\epsilon)$. We see
that the capacitance $C_q$ is given by an integral of the density
of states $\rho(\epsilon)$ over the energy window from $-e\Vexc$
to $+e\Vexc$. Under optimal operating conditions, where $\phi=0$,
a peak in $\rho(\epsilon)$ is centered around $\epsilon=0$, such
that the current becomes independent of $\Vexc$. This is clearly
visible on the diamonds in Fig.\ \ref{fig-2DplotACcurrent}.

In the following we restrict our considerations to excitation
drives that exactly compensate the level spacing, i.\ e.,
$2e\Vexc=\Delta$. In this case, one of the peaks in
$\rho(\epsilon)$ is always fully integrated over regardless of
$\phi$, and the capacitance $C_q$ becomes independent of $\phi$
and $D$\cite{Feve2007}. In that case we obtain the simple result
$C_q=e^2/\Delta$. For a square excitation drive voltage, the
time-dependent average current is
\begin{equation}
\langle \hat{I}(t) \rangle=\frac{e}{\tau}\frac{e^{-t/\tau}}{1+e^{-T/2\tau}}
\label{eq:averageI_t-dep}
\end{equation}
for $0 \leq t \leq T/2$. Clearly, this is an exponentially decaying current with a characteristic $RC$ time given by the escape time $\tau=R_q C_q$, that is related to the QPC transmission $D$ as \cite{Buttiker2009}
\begin{equation}
\tau  \simeq   \frac{h}{ \Delta} \times \left(\frac{1}{D}-\frac{1}{2}\right) \simeq \frac{\tau_o}{D} \;,\,\quad D\ll 1,
\label{eq-tauvsD}
\end{equation}
where we recall that $\tau_o=h/\Delta$ is the time it takes an electron to complete one round inside the mesoscopic capacitor. Integrating next the current over one half-period of the driving we find the average transferred charge per half-period
\begin{equation}
Q^t=2\Vexc C_q \times \tanh(\frac{1}{4f_d \tau}) = e \times \tanh(\frac{1}{4f_d \tau})\,.
\label{eq-Qt}
\end{equation}

\begin{figure}%[!htph]
\centering\includegraphics[width=0.47\textwidth]{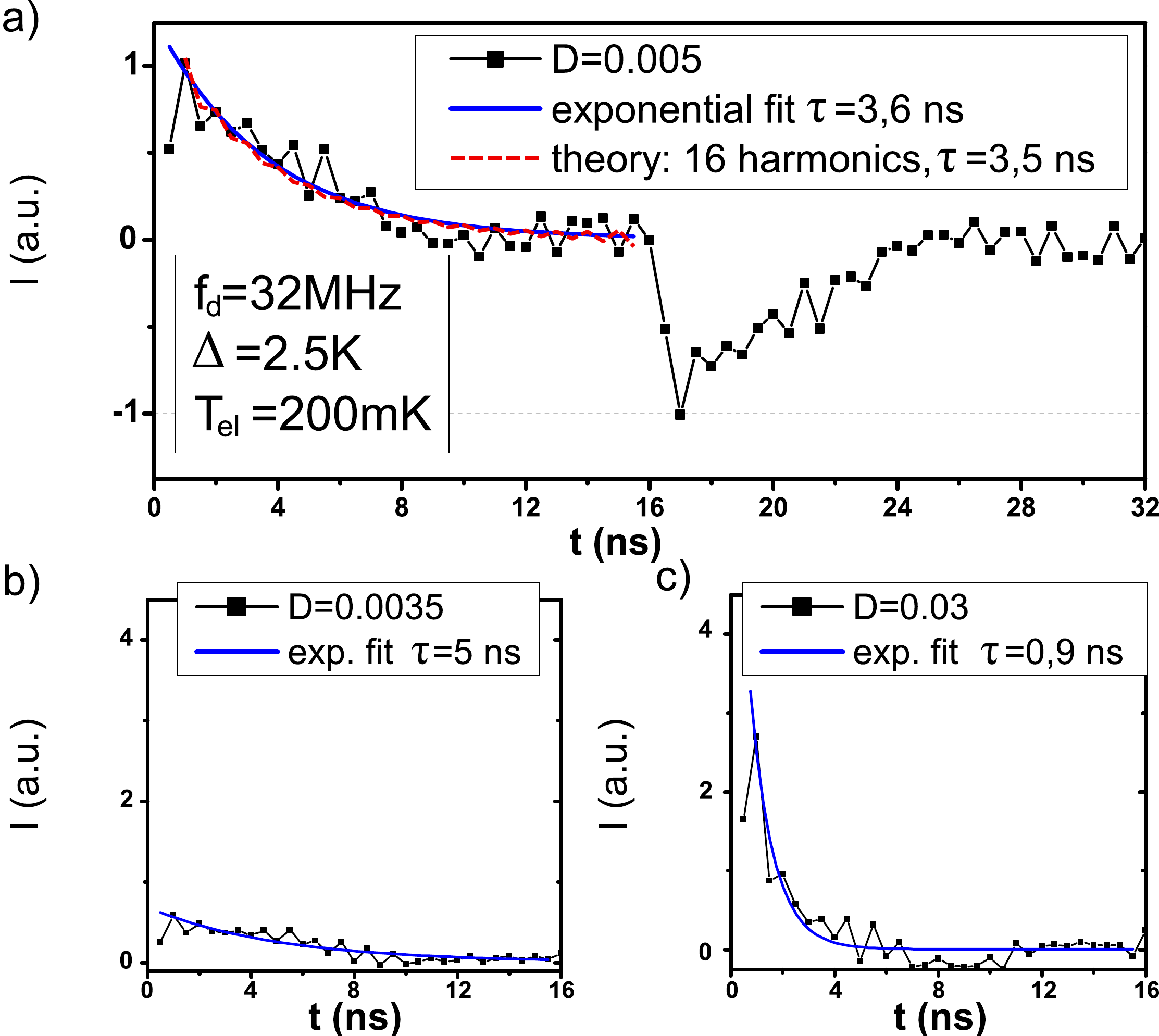}
\caption{(color online). Time dependence of the average AC current
$\langle \hat{I}(t) \rangle$ measured in sample $B$ for different
values of the QPC transmission $D$. The level spacing and electron
temperature of sample $B$ were $\Delta=2.5$ K and
$T_\mathrm{el}\approx200$ mK, respectively. The driving frequency
was $f_d=32$ MHz. Experimental data is shown with symbols, while
the blue lines are exponential fits from which the escape time
$\tau$ is extracted. The green dashed line in panel {\bf a)}
corresponds to scattering theory calculations with 16 harmonics
included. QPC transmission and extracted escape times are {\bf a)}
$D=0.005$, $\tau\approx3.6$ ns, {\bf b)} $D=0.0035$,
$\tau\approx5$ ns, and {\bf c)} $D=0.03$, $\tau\approx0.9$ ns. }
\label{fig-Ivstime}
\end{figure}

The exponential decay of the current described by Eq.\
(\ref{eq:averageI_t-dep}) was observed experimentally in Refs.\
\onlinecite{Feve2007} and \onlinecite{Mahe2008}. Typical
experimental results are shown in Fig.\ \ref{fig-Ivstime} for a
sample that we label as sample
$B$\cite{Gabelli2006,Feve2007,Mahe2008}. For this sample, the
level spacing was $\Delta=2.5$ K and the electron temperature
$T_\mathrm{el}\approx200$ mK, while the experiment was carried out
at a driving frequency of $f_d=32$ MHz.

The current is well approximated by an exponential decay, allowing
us to extract the escape time $\tau$, which is tunable by the QPC
gate voltage $V_g$ and thus the QPC transparency $D$. For
sufficiently large QPC transmissions, the integral of the current
becomes constant (see also data in Ref.\ \onlinecite{Mahe2008}),
demonstrating the quantization of the \emph{average} transferred
charge per half-period $Q^t=e$, so long as $f_d \tau \ll 1$ and
thus $\tanh(\frac{1}{4f_d \tau})\simeq 1$ in Eq.\ (\ref{eq-Qt}).
For low transmissions, $\langle \hat{I}(t) \rangle$ still exhibits
an exponential decay with an escape time $\tau$ that is comparable
to the half-period $T/2$. This indicates that the $RC$ circuit
description of the single electron emitter is still valid even for
$f_d \tau \simeq 1$. The $RC$ circuit description of the single
electron emitter also allows us to extract $\tau=R_q C_q$ from
measurements of the phase of the first harmonic $I_{\Omega}$ and
$Q^t$ from both modulus and phase measurement of $I_{\Omega}$ for
arbitrary values of the transmission $D$.

\begin{figure}%[!htph]
\centering\includegraphics[width=0.48\textwidth]{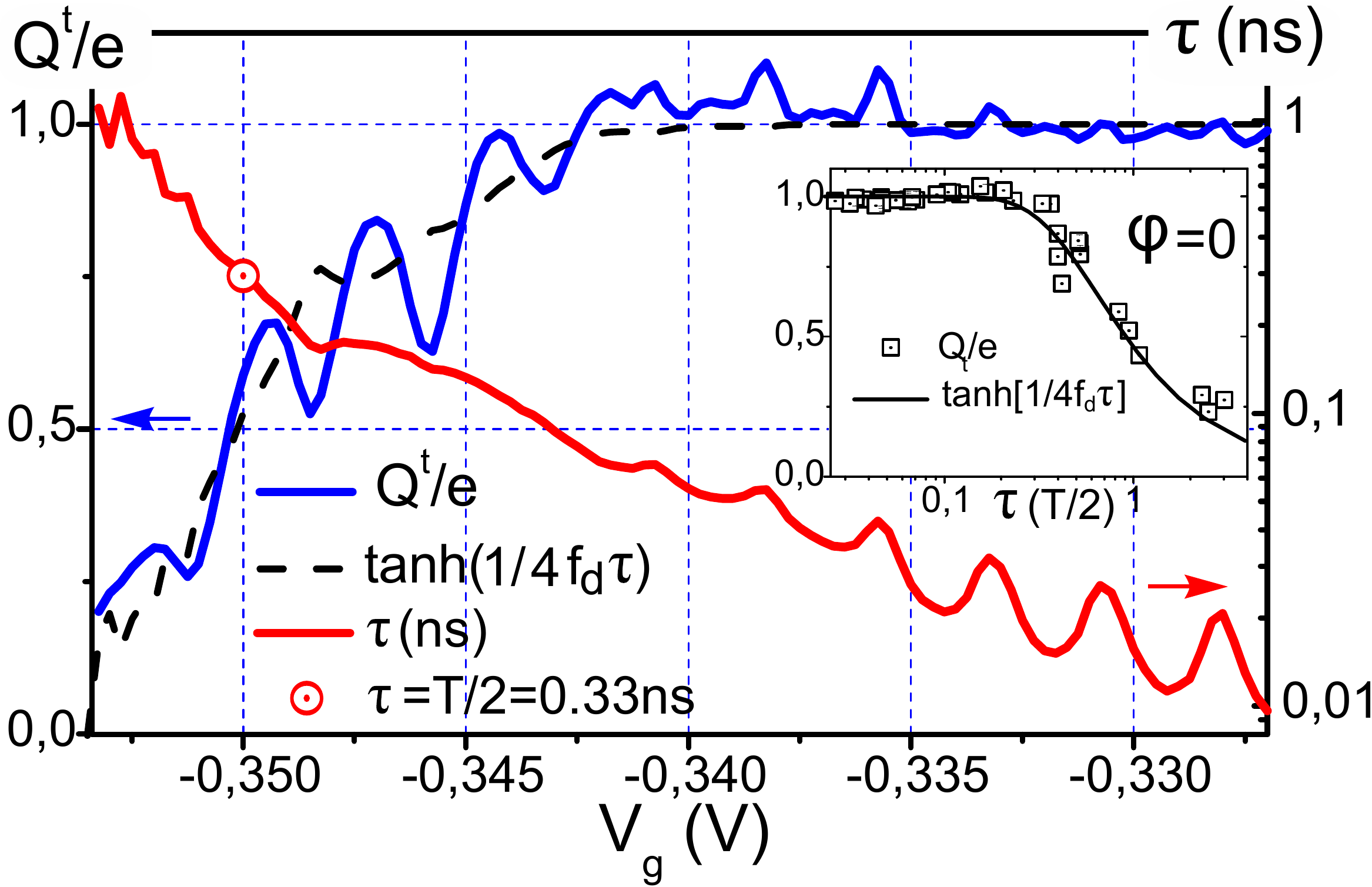}
\caption{(color online). Measured average transferred charge per
half-period $Q^t$ (blue line) and escape time $\tau$ (red line)
for sample $A$ as functions of $V_g$. Parameters are $\Delta=4.2$
K, $T_\mathrm{el}\approx60$ mK, and $f_d=1.5$ GHz. The dashed line
corresponds to $Q^t$ determined from the measured $\tau$ using
Eq.\ (\ref{eq-Qt}). The dotted circle corresponds to the point at
which $\tau=T/2$. Inset: average transferred charge for optimal
operating conditions ($\phi \approx 0$) as a function of the
escape time $\tau$ in units of $T/2$. } \label{fig-expQtauvsVg}
\end{figure}

Using measurements of the modulus (see Figure
\ref{fig-2DplotACcurrent}) and phase of $I_{\Omega}$ for sample A,
the $Q^t$ and $\tau$ dependence on the QPC gate voltage $V_g$ at
excitation amplitude $2e\Vexc=\Delta$ are plotted in Fig.\
\ref{fig-expQtauvsVg}. As $V_g$ is swept from large negative
voltages towards zero, the transmission $D$ increases, while the
escape time $\tau$ decreases over two orders of magnitude. The
blue line corresponds to the measurement data of $Q^t$ while the
black dashed line expresses $Q^t$ in terms of $\tau$ according to
Eq.\ (\ref{eq-Qt}). For sufficiently short escape times $\tau$,
$Q^t$ becomes quantized and equal to the electron charge $e$
(corresponding to the quantization of the modulus of the first
harmonic in units of $2ef$ in Fig.\ \ref{fig-2DplotACcurrent}).
Small residual oscillations around $Q^t=e$ stemming from the
capacitive coupling between the QPC gates and the cavity can be
seen. When $V_g$ is swept, the quantum dot goes periodically from
optimal conditions $\phi=0$ ($\epsilon_0=0$) to $\phi=\pi$
($\epsilon_0= \Delta/2)$. In the first case, $Q^t$ is quantized,
i.\ e.\ $Q^t=e$, whereas in the second case, $Q^t$ becomes
extremely sensitive to the exact value of the excitation amplitude
and $Q^t \approx e$ (slightly above $e$ in Fig.\
\ref{fig-expQtauvsVg}) explaining the oscillations. The
quantization of $Q^t$ can be checked on the inset of Fig.\
\ref{fig-expQtauvsVg}, showing $Q^t$ as a function of the escape
time $\tau$ for conditions close to the optimal value of $\phi
\approx 0$. For short escape times, all data agree with $Q^t=e$
within error bars (size of squares). In this regime, electrons and
holes appear to be systematically emitted with the uncertainty in
the emission time determined by $\tau$. In contrast, when the
escape time becomes comparable to half a period, the expected
quantization is lost, $Q^t<e$, reflecting that single charges are
not deterministically and periodically emitted from the quantum
dot. Furthermore, $Q^t/e$ (squares) is well described by the
expected $\tanh{(1/4f_d \tau)}$ dependence (continuous black
line). Finally, we note that the Floquet scattering matrix theory
predicts rapid oscillations in the current\cite{Moskalets2008} on
time scales on the order of $\tau_o=h/\Delta$ which is much
smaller than the periods considered here. The observation of such
oscillations, however, is not yet within experimental reach, as
the time scale $\tau_o\approx 10$ ps is still well below the
measurement resolution of our experiment.

As we have seen in this section, the measurements of the average
AC current suggest that the mesoscopic capacitor in certain
parameter regimes acts as a controllable single electron source.
Average measurements only, however, do not reveal possible
fluctuations of the quantized current and to this end we need to
consider the noise properties of the mesoscopic capacitor. As
already discussed, the noise spectrum can be calculated
numerically using the Floquet scattering matrix theory. In order
to understand in detail the noise properties of the mesoscopic
capacitor we first analyze a simple semi-classical model of the
charge transport. The description of the mesoscopic capacitor in
terms of an AC driven $RC$ circuit forms the basis of the
semi-classical model described in the following section.

\section{Semi-classical model}
\label{sec:heur_mod}

As mentioned above, the noise of the mesoscopic capacitor can be
calculated numerically using the Floquet scattering theory.
However, except for certain limiting cases, it is generally
difficult to obtain analytic results from which one could hope to
develop a deeper understanding of the noise properties. In the
adiabatic regime, where the period of the driving is much longer
than any other time scale, the noise spectrum has been found
analytically\cite{Moskalets2002,Moskalets2007,Moskalets2008}. In contrast, in the situation that we consider here, where the
driving potential is highly non-adiabatic, a perturbative
expansion in the driving frequency is not possible. Nevertheless,
both the experimental and the numerical results obtained from the
Floquet scattering approach suggest that it may be possible to
calculate analytically the noise spectrum using a conceptually
simple semi-classical model. The model that we now describe was
first suggested by Mah\'e \textit{et al.}\cite{Mahe2010} and later
investigated theoretically by Albert \textit{et
al.}\cite{Albert2010}. In the following we discuss the semi-classical model from
which we derive the finite-frequency noise spectrum and provide a
thorough comparison between analytic, numerical, and experimental
results.

The semi-classical model can be explained by considering again
Fig.\ \ref{fig-capameso}b. The model assumes that the quantum dot
can emit at most one electron and one hole per period and time is
discretized in units of $\tau_o$; the time is takes an electron to
complete one round inside the mesoscopic capacitor. In the
emission phase \textcircled{\footnotesize{1}} the probability in
each time step for an electron to escape is equal to the
transparency of the QPC, namely $D$. Additionally, since the
amplitude of the periodic driving is on the order of the level
spacing $\Delta$, higher-lying states can safely be neglected and
maximally one electron can escape the mesoscopic capacitor as
re-filling is not possible. Similarly, in the absorption phase
\textcircled{\footnotesize{2}} the probability of emitting a hole
in each time step is $D$. This semi-classical model can be
theoretically formulated as a master equation in discrete time for
the probability of the mesoscopic capacitor to be occupied by an
electron. Setting the electron charge $e=1$ in the following, this
probability is equal to the average (additional) charge of the
mesoscopic capacitor $\langle Q\rangle$, where $Q=0,1$.

The master equation determines the evolution of the average charge after one time step and takes the form \cite{Mahe2010,Albert2010}
\begin{equation}\label{master_eq}
  \langle Q(t_{k+1,\ell})\rangle= \left\{\begin{array}{ll}
  (1-D)\langle Q(t_{k,\ell})\rangle & \textcircled{\footnotesize{1}}   \\
  \\
  D[1\!-\!\langle Q(t_{k,\ell})\rangle]\!+\!\langle Q(t_{k,\ell})\rangle & \textcircled{\footnotesize{2}}
  \end{array}\right. ,
\end{equation}
where we have used that $1-\langle Q\rangle$ is the probability for the mesoscopic capacitor to be empty and $t=t_{k,\ell}$ denotes time at the $k$'th time step during the $\ell$'th period. The emission (absorption) phase $\textcircled{\footnotesize{1}}$ ($\textcircled{\footnotesize{2}}$) corresponds to $k=1,2\ldots,K$ ($K+1,K+2,\ldots,2K$), where $K$ is the number of time steps in the absorption and emission phases, respectively, each of duration $T/2$.

Experimentally, the noise measurement frequency $\omega$ was
roughly equal to the driving frequency $\Omega=2\pi f_d$ and both
were much smaller than the inverse round trip time $\tau_o$, i.\
e.\ $2\pi/\omega \simeq 2\pi/\Omega \simeq 60\,\tau_o$
\cite{Mahe2010}. This allows us eventually to consider the
continuous-time limit of the model, where the step size $\tau_o$
becomes irrelevant and drops out of the problem. Interestingly,
the physics of the system is then governed by the single
dimensionless ratio $T/\tau$ of the period $T$ over the escape (or
correlation) time
\begin{equation}
\tau\equiv\frac{\tau_o}{\ln[1/(1-D)]}.
\label{eq:tau}
\end{equation}
In the limit $D\ll 1$, we recover the expression in Eq.\ (\ref{eq-tauvsD}). At the end of this section, we discuss the physical meaning of the correlation time.

The master equation can be understood by considering the average
current as it was calculated using Floquet scattering theory in
Ref.\ \onlinecite{Moskalets2008}. For the square-shaped driving
considered in this work, the current consists of one step-like
term with time step $\tau_o$ contained in an exponential envelope
function and one oscillatory part with period $\tau_o$. The latter
corresponds to the rapid oscillations of the current mentioned in
the previous section. These oscillations are due to quantum
interferences between different orbits in the mesoscopic capacitor
and they vanish at high temperatures. Still, at arbitrary
temperature only the first step-like term survives after
integration over the time step $\tau_o$, which leads to the master
equation. The semi-classical description reflects that the
oscillations due to quantum interferences are irrelevant for the
average current on a time scale that is larger than~$\tau_o$.

At this point we have not provided a detailed derivation of the
model which would require us to compare not only average current
but also noise and higher-order correlations with the full quantum
theory. This still remains a challenging and open task and for now
we simply rely on the excellent agreement with experimental data
as we demonstrate in the following. Obviously, the semi-classical
description cannot be correct under all operating conditions and
already now we can anticipate situations where the model will
differ from the full Floquet scattering theory: for example, as
the noise measurement frequency approaches the internal frequency
of the quantum dot, namely the frequency $h/\Delta=1/\tau_o$
associated with the level spacing $\Delta$, we expect that the
semi-classical model will no longer be valid. The model also
neglects the possibility of emitting two electrons within the same
period and will therefore not apply to situations, where a level
of the quantum dot is in resonance with the Fermi level of the 2D
electron gas, i.\ e.\ for $\phi=\pi$. We make a detailed
comparison between the semi-classical model and numerics in the
following section.

Before turning to calculations of the noise spectrum, we consider
the average charge $\langle Q\rangle$ in the mesoscopic capacitor.
In Ref.\ \onlinecite{Albert2010} the average charge was used to
obtain the average current flowing out of the mesoscopic capacitor
$\langle I(t)\rangle\equiv -\langle
\dot{Q}(t)\rangle\simeq[\langle Q(t)\rangle-\langle
Q(t+\tau_o)\rangle]/\tau_o$. Solving Eq.\ (\ref{master_eq}) for
the average charge $\langle Q\rangle$ we readily find
\begin{equation}\label{qt2}
  \langle Q(t_{k,\ell})\rangle= \left\{\begin{array}{ll}
  \displaystyle \alpha_\ell\, e^{-(t_{k,\ell}-\ell T)/\tau} & \textcircled{\footnotesize{1}}     \\
  \\
  \displaystyle 1-\beta_\ell\, e^{-(t_{k,\ell}-[\ell +\frac{1}{2}]T)/\tau} & \textcircled{\footnotesize{2}}
\end{array}\right. .
\end{equation}
Here, we have defined $\varepsilon=e^{-T/2\tau}$,
$\alpha_\ell=1/(1+\varepsilon)+\theta \varepsilon^{2\ell}$ and
$\beta_\ell=1/(1+\varepsilon)-\theta \varepsilon^{2\ell-1}$ with
$\theta$ depending on the initial conditions at the time when the
periodic driving is turned on. The correlation time $\tau$
determines the time scale over which the system loses memory about
the initial conditions encoded in $\theta$ and $\langle Q\rangle$
becomes periodic in time. We notice the close relation of the
above expression with the average current of an $RC$ circuit.
Importantly, the model reproduces the expressions in Eqs.\
(\ref{eq:averageI_t-dep}) and (\ref{eq-Qt}) for the time-dependent
average current and the average transferred charge per
half-period, respectively, obtained from the Floquet scattering
theory.

\section{Noise spectrum}
\label{sec:noise_see} We are now ready to discuss the noise
properties of the mesoscopic capacitor. As we will see, the
finite-frequency noise spectrum allows us to characterize the
mesoscopic capacitor as a single electron source as well as to
determine the optimal operating conditions of the device. Before
presenting any detailed calculations we discuss the two primary
sources of noise.

\subsection{Sources of noise}

\begin{figure}
  \centering\includegraphics[width=0.47\textwidth]{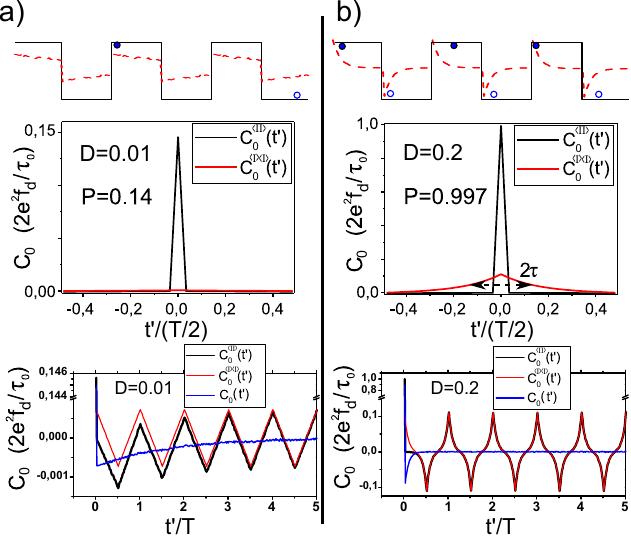}
  \caption{(color online). Current correlators within the semi-classical description. {\bf a)} Shot noise limit: for small escape probabilities $D$, charges are not systematically emitted within each half-period: the emission probability per half-period $P=Q^t/e$ is small, and the current fluctuations correlator is essentially given by the Dirac peak in $C_{0}^{ \langle I I \rangle}(t')$. The correlators $C_{0}^{ \langle I I \rangle}(t')$, $C_{0}^{ \langle I\rangle \langle I \rangle}(t')$ and $C_0(t')$ for long time $t'$ are shown in the lower panel (with black, red, and blue lines, respectively): $C_0(t')$ is negative for $t'\lesssim 3T$, reflecting the anti-bunching of emitted charges. {\bf b)} Phase noise limit: for sufficiently large escape probabilities (here, $D=0.2$), charges are systematically emitted ($P\simeq 1$), and $C_{0}^{ \langle I\rangle \langle I \rangle}(t')$ consists of a peak with a finite width given by the escape time $\tau$. In this case, $C_0(t')$ is negative in a smaller range of times $t'\lesssim T/2$.
  }
  \label{fig-heurC1C2}
\end{figure}

In the semi-classical model, the mesoscopic capacitor emits at
most one electron and one hole per period. Still, depending on the
ratio between the escape time $\tau$ and the period $T$, the
source may fail to emit. We quantify the emission probability per
half-period by the ratio
\begin{equation}
  P=Q^t /e,
  \label{eq-PQ}
\end{equation}
having recalled that $Q^t$ is the average transferred charge per cycle, see Eq.\ (\ref{eq-Qt}). We refer to the noise associated with such cycle missing events, where the mesoscopic capacitor fails to emit, as \emph{shot noise}. However, even when the emission probability is close to unity and the mesoscopic capacitor emits an electron and a hole in almost every cycle, there are still fluctuations in the current associated with the random emission times within a period. This source of noise is referred to as \emph{phase noise}. The two uppermost panels of Fig.\ \ref{fig-heurC1C2} illustrate the main sources of noise by showing typical realizations of the current, where emissions of electrons (holes) are shown with filled (empty) circles, on top of the average current. The uppermost panel of Fig. \ref{fig-heurC1C2}a illustrates shot noise, while the uppermost panel of Fig.\ \ref{fig-heurC1C2}b corresponds to phase noise.

From the definition of the current-current correlation function in Eq.\ (\ref{eq:currcurr_corr}) we can immediately write
\begin{equation}
C(t,t')=\langle I(t) I(t+t')\rangle-\langle I(t)\rangle \langle I(t+t')\rangle,
\end{equation}
where $I(t)$ is no longer a quantum mechanical operator, since we are considering a semi-classical description. As already mentioned, the system is not translational invariant in time due to the periodic gate voltage modulations. The correlation function therefore does not only depend on the time difference $t'$, but also on the absolute time $t$. Experimentally, the correlation function is averaged over the absolute time $t$ and the time-average correlation function is then
\begin{equation}
C_0(t')=C_{0}^{ \langle I I \rangle}(t')-C_{0}^{ \langle I\rangle \langle I \rangle}(t'),
\end{equation}
where $C_{0}^{ \langle I I \rangle}(t')=\overline{\langle I(t) I(t+t')\rangle}^{\,t}$ and $C_{0}^{ \langle I\rangle \langle I \rangle}(t')=\overline{\langle I(t)\rangle \langle I(t+t')\rangle}^{\,t}$. The time-average correlation function is exactly the Fourier component $C_0(t')$ entering Eq.\ (\ref{eq:Four_Ser}) and the corresponding noise spectrum is consequently given by Eq. (\ref{eq-noisespectrum}).

Figure \ref{fig-heurC1C2} shows numerical calculations of the correlation functions in the two limiting cases. The results were obtained in numerical simulations of the stochastic process defined by the semi-classical model. The left panels correspond to the shot noise regime, where the escape time $\tau$ is much larger than the period $T$, whereas the right panels show results for the phase noise regime. Depending on the escape probability $D$ [or equivalently the ratio between the escape time and the period, see Eq.\ (\ref{eq:tau})], the contributions from $C_{0}^{ \langle I I \rangle}(t')$ and $C_{0}^{ \langle I\rangle \langle I \rangle}(t')$ vary. The two central panels of Fig.\ \ref{fig-heurC1C2} show the correlation functions $C_{0}^{ \langle I I \rangle}(t')$ and $C_{0}^{ \langle I\rangle \langle I \rangle}(t')$ for short times $t'\ll T$. In both regimes, $C_{0}^{ \langle I I \rangle}(t')$ contains a Dirac peak at $t' = 0$. Indeed, since maximally one charge is emitted per half-period, the short-time correlations vanish. In this respect, the Dirac peak is the hallmark of single particle emission. Correlations are only recovered when $t'$ is close to a multiple of the half-period, as seen in the lower panels of Fig.\ \ref{fig-heurC1C2}.

The height of the Dirac peak at $t'=0$ is proportional to the average transferred charge per half-period $Q^t$: $C_{0}^{ \langle I I \rangle} (t' = 0)$ counts the average number of peaks and dips in the instantaneous current $I(t)$ corresponding to emitted electrons and holes. The correlation function $C_{0}^{ \langle I\rangle \langle I \rangle}(t')$ is given by the autocorrelation of the exponentially decaying average current. At short times, it therefore has a peak at $t'=0$ with a finite width given by the escape time $\tau$. At times comparable to multiples of half a period, correlations are again recovered, which compensate the long-times correlations in $C_{0}^{\langle I I \rangle}(t')$. Thus, for long times $t'\gg T$, the correlation function $C(t')$ vanishes since charges emitted by the source are no longer correlated. The timescale on which $C(t')$ decays to zero depends on the transmission $D$ and is given by the escape (or correlation) time $\tau$.

\subsubsection{Shot noise}
\label{subsec:shot_noise}

For small escape probabilities $D\ll 1$, the escape time $\tau$
becomes much larger than the period $T$. The peak at $t'=0$ in
$C_{0}^{ \langle I\rangle \langle I \rangle}(t')$ correspondingly
becomes much smaller than the Dirac peak in $C_{0}^{ \langle I I
\rangle}(t')$, see Fig.\ \ref{fig-heurC1C2}a. The current
correlator $C(t')$ is thus given by a Dirac peak at $t' = 0$
combined with negative values up to at finite times on the order
of the correlation time $\tau$. The noise power spectral density
is then constant, except at zero frequency, where it vanishes
because the integrals over $C_{0}^{ \langle I I \rangle}(t')$ and
$C_{0}^{ \langle I\rangle \langle I \rangle}(t')$ cancel each
other. In this case, the source randomly emits charges and the
charge fluctuations are similar to shot noise.

The negative values of $C(t')$ at finite times reflect the anti-bunching behavior of emitted charges: at low transmission probabilities $D$, the anti-bunching extends over a large range of times, $t'\simeq\tau$, see Fig.\ \ref{fig-heurC1C2}a (lower panel). Even in the shot noise regime, a hole must be emitted after the emission of an electron before a second electron can be emitted. Now, approximating $C_{0}^{ \langle I\rangle \langle I \rangle}(t')\simeq 0$, see Fig.\ \ref{fig-heurC1C2}a, and writing $C_{0}^{ \langle I I \rangle}(t') = (2 f_d Q^t /e) \delta (t')$, we find\cite{Mahe2010, Albert2010}
\begin{equation}\label{eq_noise_shot}
  S_{\rm shot}(\omega)=4e^2 f_d \times P=\frac{e^2}{\tau} ,
\end{equation}
where $P$ the emission probability per half-period, Eq.\ (\ref{eq-PQ}). Interestingly, this expression is identical to the usual shot noise formula $S=2eI_p$, where the usual charge current has to be replaced by the particle current, $I_p$, given by the sum of the average number of electrons emitted in the first half period and holes in the second one, times the product of the electric charge with the drive frequency:  $I_p = 2ef_d \times P$.

\subsubsection{Phase noise}
\label{subsec:phas_noise}

In the phase noise regime, the escape probability $D$ is so high that charges are emitted in nearly every cycle and the emission probability $P$ is close to unity. The time-dependent average current then consists of well-defined exponential decays with a decay time given by the escape time $\tau\ll T/2$: $\langle I(t) \rangle= \pm (e/\tau) e^{- t/\tau}$. Here the different signs correspond to the emission of electrons or holes. In this case, we find a simple expression for the correlation function
\begin{equation}
  C_{0}^{ \langle I\rangle \langle I \rangle}(t')=\frac{e^2 f_d}{\tau}e^{-|t'|/\tau}\,.
\end{equation}
and the noise is then given by\cite{Mahe2010, Albert2010}
\begin{equation}\label{eq_noise_jitter}
  S_{\rm phase}(\omega)=4e^2 f_d \frac{\omega^2 \tau^2}{1+\omega^2 \tau^2}\,.
\end{equation}
Even if charges are systematically emitted each period, we find a finite value of the noise which depends only on the escape time $\tau$. This noise is due to the uncertainty in the emission time of charges within a period, and is thus referred to as phase noise. The phase noise is an intrinsically high-frequency noise and it is the signature of single charge emission: when the source periodically emits single charges, the noise reduces to the value of the phase noise determined only by the temporal extension $\tau$ of the emitted wave packets.

\begin{figure*}
  \begin{center}
    \includegraphics[width=0.8\textwidth]{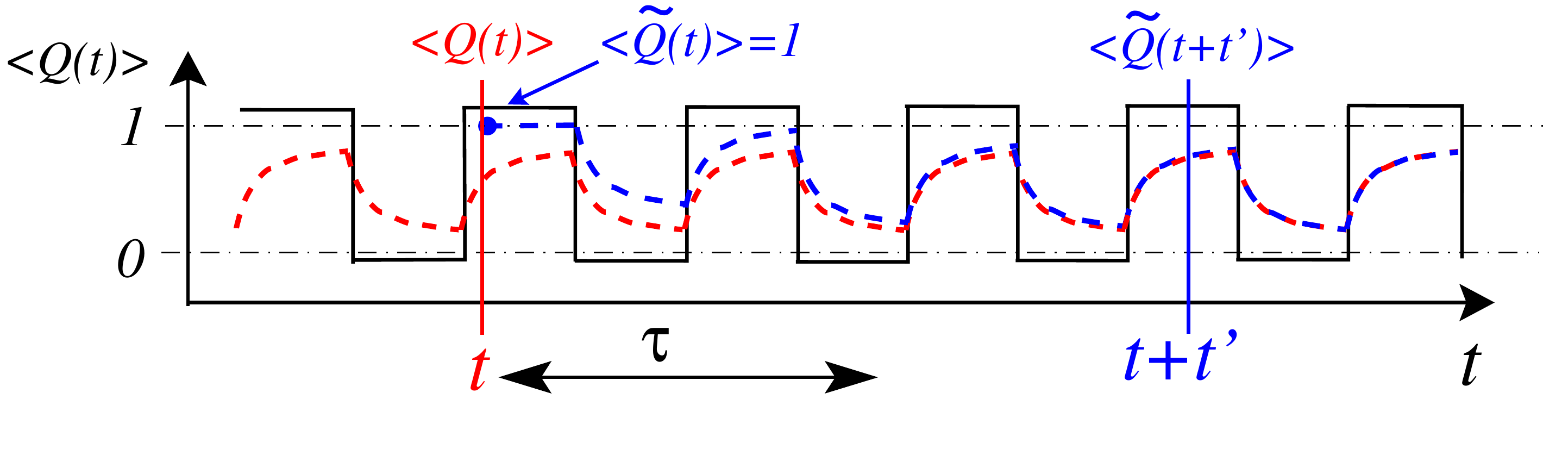}
    \caption{(color online). Schematics of periodic driving and (conditional) charge occupations. The mean occupation of the mesoscopic capacitor $\langle Q(t)\rangle$ is shown in red. The mean occupation is equal to the probability of the mesoscopic capacitor to be occupied with one (additional) electron. The blue curve shows the \emph{conditional} mean occupation $\langle \widetilde Q(t+t')\rangle$. This is the probability that the mesoscopic capacitor is charged with an (additional) electron at time $t+t'$, given that it was charged at time $t$, such that $\langle \widetilde Q(t)\rangle=1$. Correlations decay on a time scale set by the correlation time $\tau$, implying that $\langle \widetilde{Q}(t+t')\rangle\simeq  \langle Q(t+t')\rangle$ for $t'\gg \tau$.}
    \label{fig:calcnoise}
  \end{center}
\end{figure*}

\subsection{Analytic expression}
\label{subsec:ana_form}

We now present an analytic calculation of the noise power spectrum which covers both limiting cases as well as the intermediate regime \cite{Albert2010}. To this end, it is useful to consider the charge correlation function $C_{Q}(t,t')=\langle Q(t)Q(t')\rangle-\langle Q(t)\rangle\langle Q(t')\rangle$ together with the relation $\mathcal{P}_0(\omega)\simeq \omega^2 \mathcal{P}_Q(\omega)$. Here the definition of $\mathcal{P}_Q(\omega)$ is similar to that of $\mathcal{P}_0(\omega)$ in Eq.\ (\ref{eq-noisespectrum}), but with the current $I$ replaced by the charge $Q$. The charge correlation function can be evaluated following the schematic illustration in Fig.\ \ref{fig:calcnoise}. We first note that $\langle Q(t)Q(t+t')\rangle$ is the joint probability for the capacitor to be charged with one electron both at time $t$ and at time $t+t'$. Using conditional probabilities we then write $\langle Q(t)Q(t+t')\rangle=\langle Q(t)\rangle \langle \widetilde Q(t+t')\rangle$, where $\langle \widetilde Q(t+t')\rangle$ is the probability that the capacitor is charged with one electron at time $t+t'$ given that it is charged at time $t$. For $t'>0$, the conditional probability $\langle \widetilde Q(t+t')\rangle$ can be found by propagating forward in time the condition $\langle \widetilde Q(t)\rangle=1$ using the master equation in Eq.\ (\ref{master_eq}), see also Fig.\ \ref{fig:calcnoise}. Similar reasoning applies to the case $t'<0$. Finally, integrating over $t$, the time-averaged charge correlation function becomes $\overline{\langle\delta Q(t)\delta Q(t+t')\rangle}^{\,t}=\,\frac{\tau}{T}\,e^{-|t'|/\tau}\tanh\left(\frac{T}{4\tau}\right)$, and we immediately obtain the noise power spectrum as \cite{Albert2010}
\begin{equation}\label{pnotw}
  S(\omega)=4e^2 f_d \; \tanh\left(\frac{1}{4f_d \tau}\right)\,\frac{\omega^2\tau^2}{1+\omega^2\tau^2}\,.
\end{equation}
Interestingly, the noise power spectrum is given by the average charge emitted during the emission phase $\tanh(1/4f_d\tau)$ (the factor of 2 accounts for the additional contribution from the average charge absorbed in the absorption phase) multiplied by a Lorentzian-like frequency dependence, which accounts for the exponential decay of correlations in the time domain with time constant $\tau$. Finally, the factor $\omega^2\tau^2$ reflects that the noise spectrum becomes flat in the high-frequency limit, while the zero-frequency limit $S(0)=0$ shows that charge on average does not accumulate on the capacitor once $\langle Q\rangle$ has become periodic in time. From Eq.\ (\ref{pnotw}) it is straightforward to recover the limiting cases given by Eqs.\ (\ref{eq_noise_shot}) and (\ref{eq_noise_jitter}), but the analytic result above also accounts for the intermediate regime where the escape time $\tau$ is comparable to the period $T$.

\subsection{Detailed comparison}
\label{comparison}

\begin{figure}%[!htph]
\centering\includegraphics[width=0.48\textwidth]{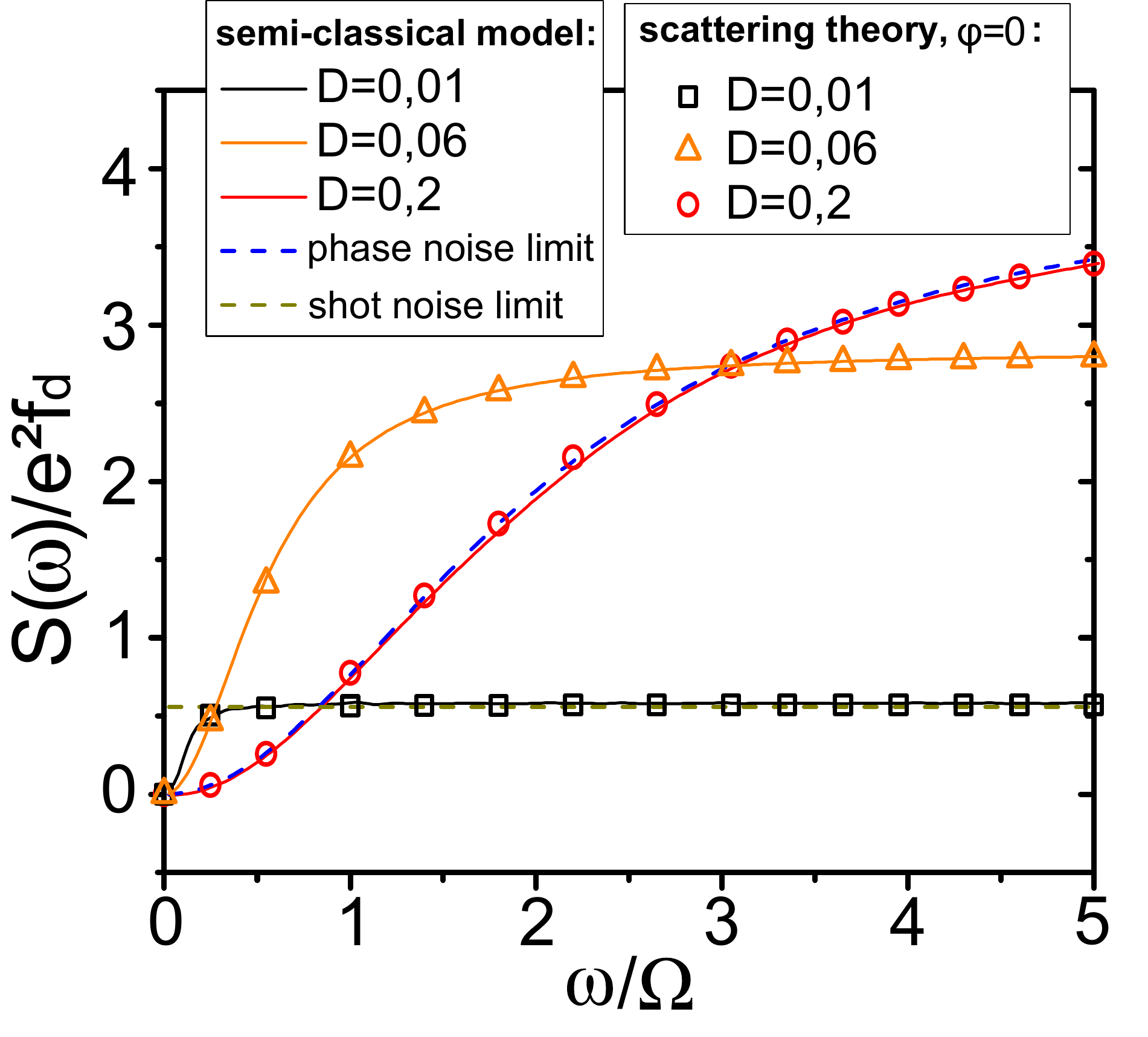}
\caption{(color online). Finite-frequency noise $S(\omega)$ close
to the driving frequency, $\omega \sim \Omega$. Symbols:
scattering theory ($\phi=0$); continuous lines: semi-classical
model; dashed lines: limits given by Eqs.\ (\ref{eq_noise_shot})
and (\ref{eq_noise_jitter}). Parameters are $\Delta=4K$ and
$f_d=1.5$ GHz.} \label{fig-BvsomHS}
\end{figure}

We are now ready to carry out a careful comparison of experimental and theoretical results. In Fig.\ \ref{fig-BvsomHS} we first compare results for the noise spectrum obtained from the full Floquet scattering theory and the semi-classical model. We focus here on the experimentally relevant regime with $\omega \approx \Omega \ll \Delta/\hbar$ together with the optimal operating conditions of the source, $\phi=0$, and consider three different values of the transmission probability $D$. The figure shows that the two complementary approaches yield results that are in excellent agreement. We observe that our numerical and analytic calculations agree well both in the shot noise limit with $D=0.01$ ($\tau\approx 3.5\:T/2$) and in the phase noise limit with $D=0.2$ ($\tau\approx 0.16\:T/2$), as well as in the intermediate cross-over regime. For $\omega\tau\gg 1$, the phase noise $S_{\rm phase}$ saturates to $4e^2 f_d$, such that $S(\omega)\simeq 4 e^2 f_d P$ as seen both for $D=0.06$ ($\tau\approx0.57\:T/2$, $P=0.71$) and $D=0.01$ ($P=0.14$).

\begin{figure}%[!htph]
\centering\includegraphics[width=0.45\textwidth]{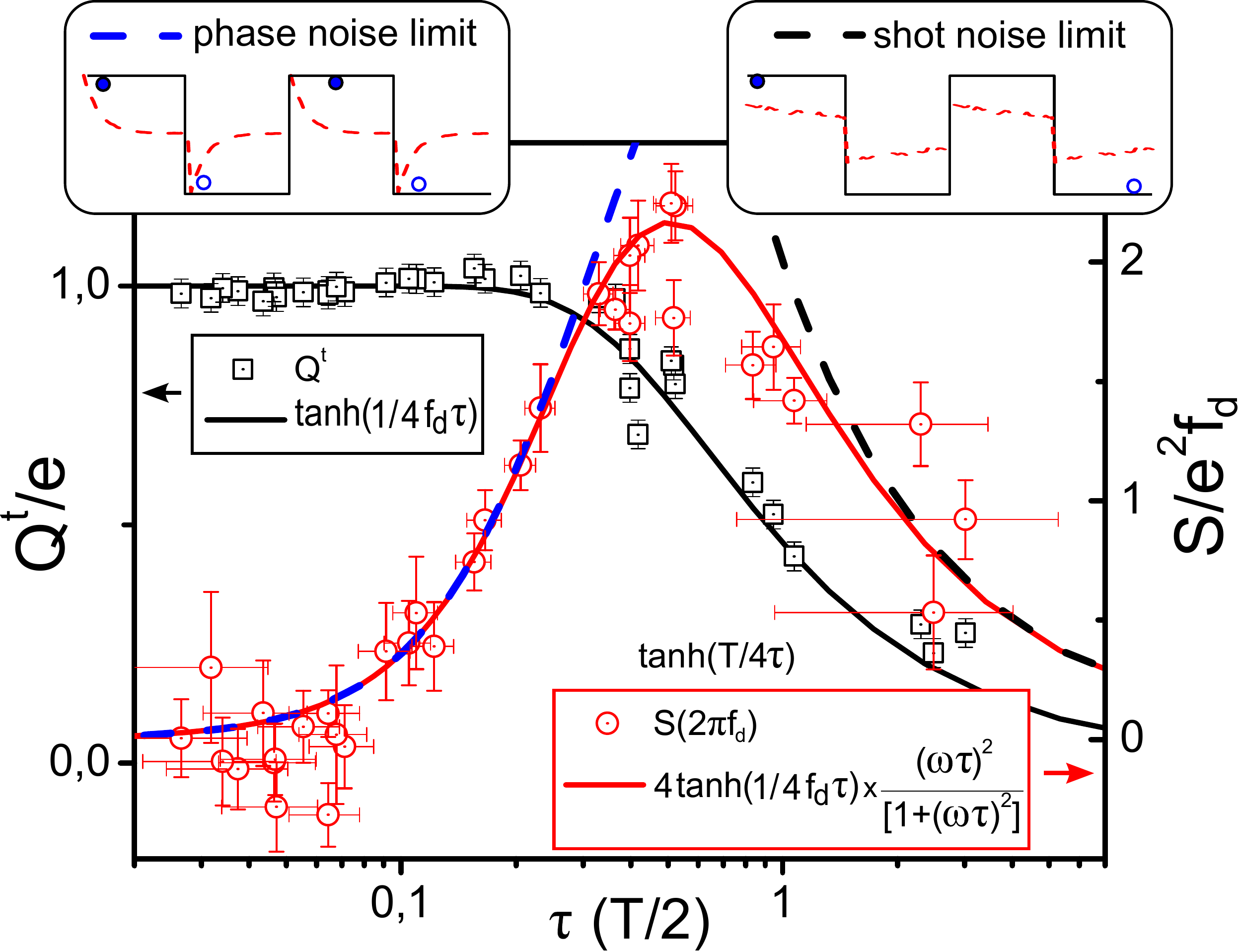}
\caption{(color online). Noise power spectrum $S(\omega=\Omega)$
as a function of the escape time $\tau$. Left axis: measured
average transferred charge per half-period $Q^t$ (black squares)
together with Eq.\ (\ref{eq-Qt}) (continuous black line). Right
axis: measured noise spectrum (red circles, $\phi\approx0$, sample
A) together with Eq.\ (\ref{pnotw}) (red line). Dashed lines
correspond to the two noise limits, Eqs.\ (\ref{eq_noise_shot})
and (\ref{eq_noise_jitter}), illustrated schematically in the top
panels. } \label{fig-Bvstauexp}
\end{figure}

The two theoretical approaches can also be compared with experimental data, obtained with sample A using a high sensitivity microwave noise measurement setup implemented in a dilution refrigerator \cite{Parmentier2011}. The noise was measured in a bandwidth of $1.2-1.8$ GHz centered around the driving frequency $f_d=1.5$ GHz. Figure \ref{fig-Bvstauexp} shows the dependence of the noise $S(\omega=\Omega)$ on the escape time $\tau$. The numerical and analytic results (overlapping continuous lines) are in excellent agreement with the experimental data (circles) obtained at $\phi \approx 0$. In particular, the experimental results are captured by the analytical expression in Eq.\ (\ref{pnotw}) over more than two orders of magnitude of the escape time $\tau$, going from the phase noise limit (blue dashed line), where the source exactly emits a single electron and a single hole in each cycle, to the shot noise limit (black dashed line) where particle emissions are rare and shot noise like\cite{Albert2010}. As discussed in Sec.\ \ref{sec-Idiscussion}, the average emission probability $P$ (squares) is also well captured by the expected $\tanh{1/4f_d\tau}$ dependence (continuous black line). The results shown in Fig.\ \ref{fig-Bvstauexp} demonstrate that the mesoscopic capacitor indeed behaves as an controllable on-demand single electron emitter when operated under optimal conditions.

\subsection{Universality and deviations}
\label{subsec:noi_prop}

While the semi-classical model is only valid for a restricted range of parameters, the Floquet scattering theory in contrast allows us to explore the full set of experimentally relevant operating conditions, including changes of temperature, level spacing, and measurement frequency $\omega/2\pi$. In the following, we first discuss a particular universal property of the noise under optimal operating conditions as described by the semi-classical model. Secondly, we discuss possible discrepancies between the semi-classical model and the full numerical calculations based on the Floquet scattering theory. As we will see below, the two descriptions start to differ once the system is not operated under optimal conditions, or when the noise is measured at very large frequencies $\hbar \omega \sim \Delta$.

\subsubsection{Universality}
\label{subsec:univ}

\begin{figure}%[!htph]
\centering\includegraphics[width=0.47\textwidth]{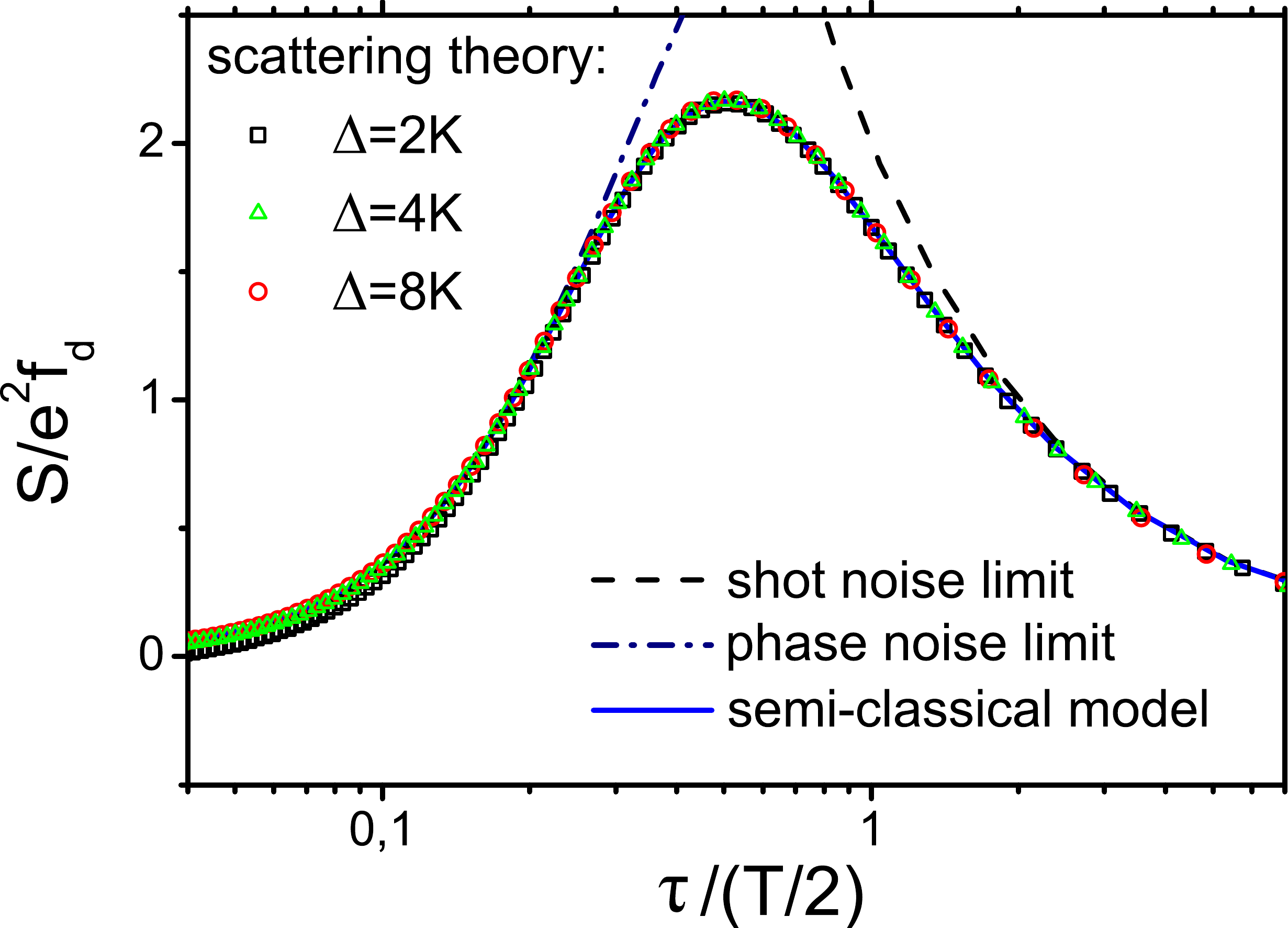}
\caption{(color online). Universality of the noise spectrum. Noise
spectrum based on the semiclassical model, Eq.\ (\ref{pnotwNorm})
(blue line), and full numerical Floquet scattering theory for
different values of the level spacing (symbols) as a function of
the escape time $\tau$ with $T_{\rm el}=100$ mK. The dashed lines
correspond to the two limits given by Eqs.\ (\ref{eq_noise_shot})
and (\ref{eq_noise_jitter}). } \label{fig-BvstauHS}
\end{figure}

Figure \ref{fig-BvsomHS} clearly illustrates that the Floquet scattering theory and the semi-classical model agree well under optimal operating conditions. Moreover, by plotting the noise power $S(\omega=\Omega)$ in units of the driving frequency $e^2f_d$ as a function of the escape time $\tau$ in units of the half-period $T/2$ one finds strong indications of a simple universal behavior of the noise power which is independent of the specific parameters of the system. Indeed, by rewriting Eq.\ (\ref{pnotw}) in term of these normalized units, we find
\begin{equation}\label{pnotwNorm}
  S(\Omega)_{[e^2f_d]}=4\tanh\left(\frac{1}{2\tau_{[T/2]}}\right)\,\frac{(\pi\tau_{[T/2]})^2}{1+(\pi\tau_{[T/2]})^2},
\end{equation}
from which this universality is evident. In particular, neither the level spacing $\Delta$ nor the temperature enter this expression. The universality of the noise spectrum is verified by our full numerical Floquet scattering theory calculations with different values of $\Delta$ as shown in Fig.\ \ref{fig-BvstauHS} (the dependence of $\tau$ on $\Delta$ is taken into account). We also checked performed calculations with varying temperatures (not shown) and found good agreement with the expression above. The universal behavior can be understood by noting that under optimal operating conditions, the noise arises from elementary charge transfer processes which only depend on the parameter $\tau$. As long as the charges are emitted sufficiently far above or below the Fermi level, these processes do not depend of the energy at which the charges are emitted.

\subsubsection{Deviations}
\label{subsubsec:devi_heur}

Deviations from the universal behavior start to appear as the QPC
transmission approaches unity, $D \simeq 1$, and the escape time
$\tau$ becomes comparable to the inverse level spacing
$\tau_o=h/\Delta$. For short escape times, Eq.\ (\ref{pnotwNorm})
predicts that the noise would vanish. However, the semi-classical
description is expected to break down as the relevant time scales
of the problem approach $\tau_o$. Small deviations between the
semi-classical model and full numerics are already visible in
Fig.\ \ref{fig-BvstauHS} for $\tau \lesssim 0.1\:T/2$. The
semi-classical model is also not expected to be valid when one of
the levels in the quantum dot is brought into resonance with the
Fermi energy during the emission cycle. In this case, the total
charge on the quantum dot is no longer quantized and the
quantization of the first harmonics of the average AC current is
lost, see Fig.\ \ref{fig-2DplotACcurrent}. Under these conditions,
the noise is expected to depend strongly on various parameters
such as temperature, the shape of the excitation drive, and the
static potential in the quantum dot.

\begin{figure}%[!htph]
\centering\includegraphics[width=0.47\textwidth]{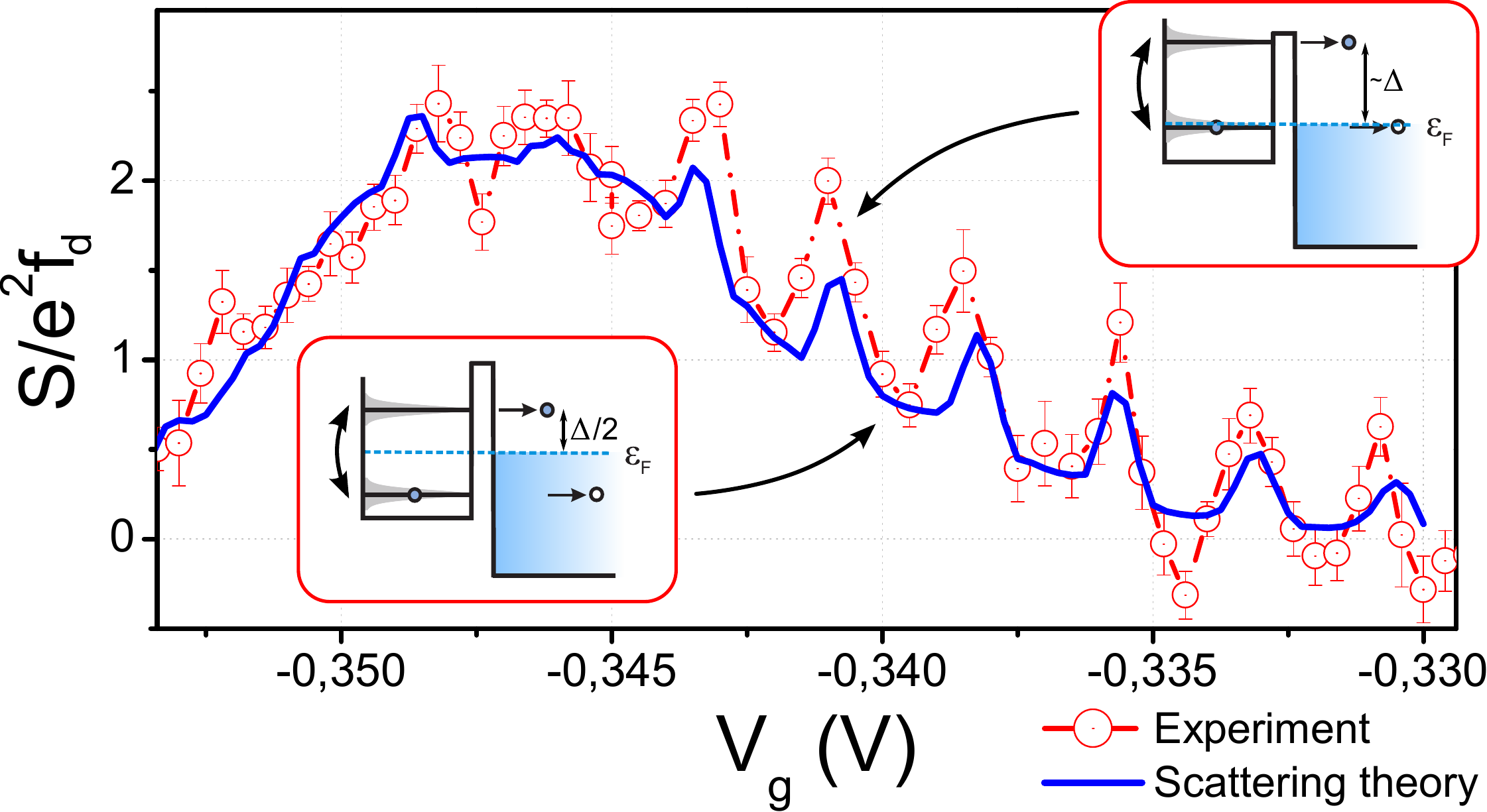}
\caption{(color online). Noise spectrum as a function of the QPC
gate voltage $V_g$. Circles: experimental data (sample $A$);
continuous line: Floquet scattering theory ($\Delta=4$ K, $T_{\rm
el}=100$ mK, 3 odd harmonics in the excitation drive). The panels
illustrate charge emission in the minima ($\phi=0$) and the maxima
($\phi=\pi$) of the noise oscillations. } \label{fig-BvsVgexp}
\end{figure}

The expected parameter dependencies are clearly visible in the noise measurements. In Fig.\ \ref{fig-BvsVgexp} we show the measured noise spectrum as a function of the QPC gate voltage $V_g$ (red circles), which simultaneously controls the QPC transmission $D$ and the levels in the quantum dot via a capacitive coupling, see also Fig.\ \ref{fig-2DplotACcurrent}. Oscillations in the noise are observed for $-0.345$~V~$\lesssim V_g \lesssim$~$-0.330$~V (0.8 $\lesssim P \lesssim 1$). The maxima of the oscillations correspond to the case $\phi=\pi$, where one of the levels is brought into resonance with the Fermi energy, while the minima occur at the optimal operating conditions $\phi=0$, see insets in Fig.\ \ref{fig-BvsVgexp}. The semi-classical model cannot account for these oscillations\cite{Mahe2010}. Instead, we have used the Floquet scattering theory and numerically evaluated Eq.\ (\ref{eq-Som3Termdiff}) as a function of $V_g$. The dependence of the QPC transmission on the gate voltage $D(V_g)$ was extracted using Eq.\ (\ref{eq-tauvsD}) in combination with the escape time $\tau$ as a function of $V_g$, Fig.\ \ref{fig-expQtauvsVg}. The AC drive $V_{\rm exc}(t)$ used in the calculations consisted of three odd harmonics, accounting for the finite bandwidth of the microwave pulse generator used in the experiments.

The numerical results (blue line) shown in Fig.\ \ref{fig-BvsVgexp} are in good agreement with the experimental data. The maxima in the noise can be understood by noting that a square-shaped AC drive with a finite number of harmonics contains fast oscillations, or ripples, which affect the energy resolution of the emitted charges. In the case $\phi=\pi$, the level brought into resonance with the Fermi energy and then oscillates rapidly with respect to the Fermi energy. This ``shaking" of a resonant level causes additional charge transfers (which also depend on the QPC transmission), leading to an increase in the noise. Due to such spurious emissions of electron--hole pairs, the source does not behave as a perfect single particle emitter under these operating conditions.

Deviations between the Floquet scattering theory and the semi-classical model also occur when the measurement frequency becomes comparable to the level spacing $\omega\simeq\Delta/\hbar$. Indeed, since the semi-classical model describes the dynamics of charge emissions with a discrete time step $\tau_o=h/\Delta$, it cannot account for dynamics on time scales shorter than $\tau$, or large measurement frequencies $\omega\simeq\Delta/\hbar$. The Floquet scattering model, on the other hand, predicts fast oscillations on time scales comparable to $\tau_o$ in the average AC current\cite{Moskalets2008}. We therefore expect strong deviations between the semi-classical model and the Floquet scattering theory in the noise spectrum at high frequencies.

\begin{figure}%[!htph]
\centering\includegraphics[width=0.43\textwidth]{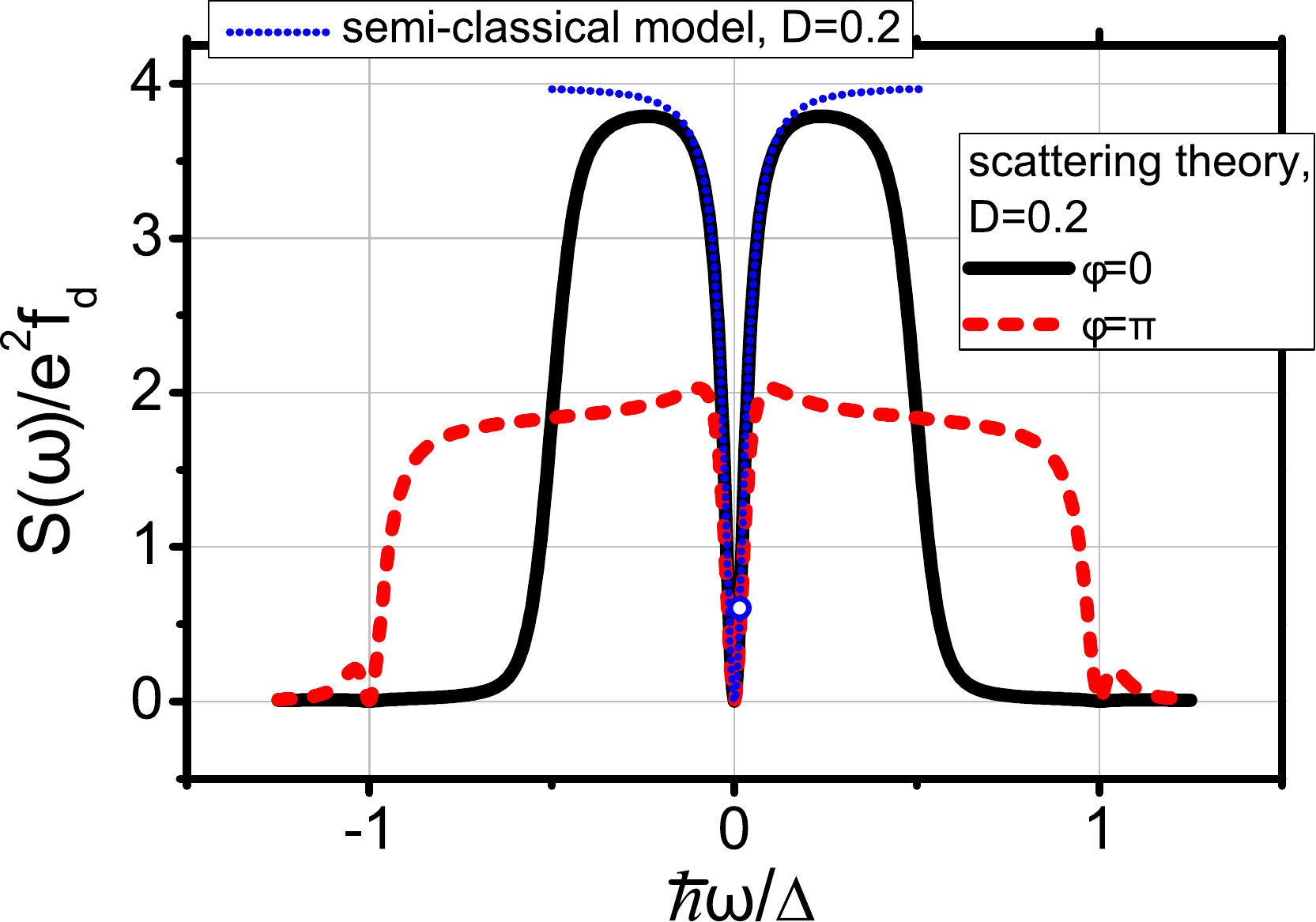}
\caption{(color online). Noise spectrum as a function of the
measurement frequency $\omega$. Blue dotted line: semi-classical
model. Since time is discretized in units of $\tau_o=h/\Delta$,
$S(\omega)$ can only be calculated up to $\hbar\omega=\Delta/2$.
Black line: Floquet scattering theory with $\phi=0$. Red dashed
line: Floquet scattering model with $\phi=\pi$. The electronic
temperature was set to $T_{\rm el}=100$ mK. The blue circle
denotes the noise at the driving frequency, $\omega=\Omega$.
Parameters are $D=0.2$ ($P\approx 1$), $\Delta=4$ K, and $f_d=1.5$
GHz. } \label{fig-BvsOmHSphi}
\end{figure}

Numerical calculation of the noise spectrum at high measurement frequencies are shown in Fig.\ \ref{fig-BvsOmHSphi} obtained from the Floquet scattering theory and the semi-classical model. We see that the two complementary approaches agree well at small frequencies for $\phi=0$. However, while the noise in the semi-classical model saturates to $S(\omega\gg\Omega)=4e^2f_d$ at high frequencies, the Floquet scattering theory, in contrast, is cut off for $\hbar\omega\approx\Delta/2$, where it drops to zero. Indeed, the electrons (holes) are emitted at an energy $\Delta/2$ above (below) the Fermi energy, corresponding to emission of radiation (or photons) at frequencies below $\Delta/2$. For $\phi=\pi$, charges can be emitted at energies up to $\Delta$ above or below the Fermi energy, and the cutoff frequency is then equal to $\Delta$ as seen in the figure. We note that the excess noise shown here indeed is symmetric in $\omega$ as expected.

\subsection{Full counting statistics}
\label{subsec:fcs}

\begin{figure}
\centering\includegraphics[width=0.47\textwidth]{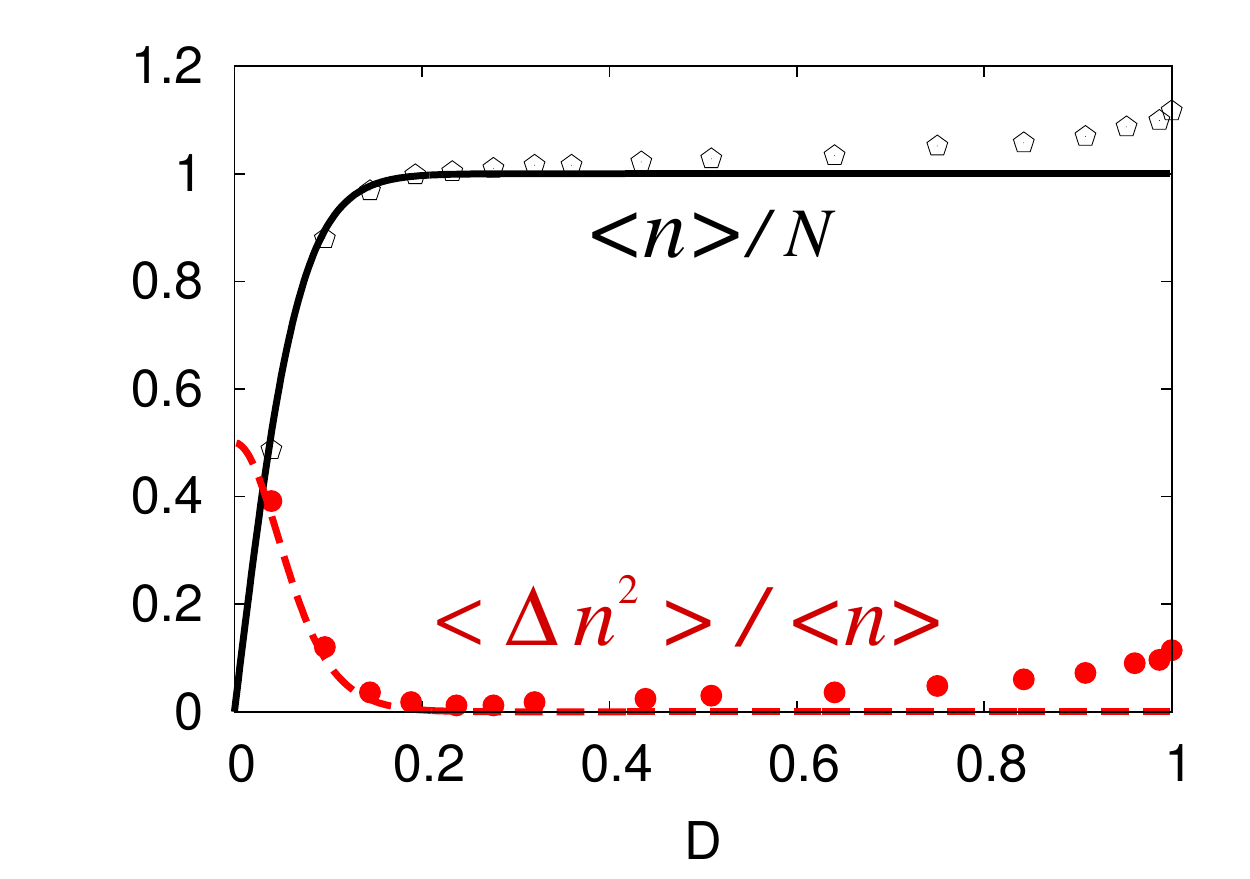}
\caption{(color online). Mean and variance of the counting
statistics of emitted electrons $P(n,N)$ after a large number of
periods $N$ as functions of the QPC transparency $D$. Full lines:
Results based on the semi-classical model. Symbols: Numerical data
obtained from the Floquet scattering theory. Numerical results
(symbols) were adapted from Ref.\ \onlinecite{Gre10}. }
\label{fig-FCS}
\end{figure}

We round off this section by discussing the counting statistics of emitted electrons. Under optimal operating conditions, the semi-classical model fully accounts for the charge dynamics of the emitter at low frequencies and it allows for tractable calculations of the counting statistics $P(n,N)$ of the number of emitted electrons $n$ during a large number of periods. In principle, we can calculate all moments (or cumulants) of the distribution\cite{Albert2010,Pistolesi2004,Battista2011}, but we focus here on the first and second cumulant, the mean $\langle n\rangle$ and the variance $\langle \Delta n^2\rangle$ with $\Delta n=n-\langle n\rangle$.
In Fig.\ \ref{fig-FCS} we show a comparison between the calculations of the first two cumulants based on the semi-classical model\cite{Albert2010} and Floquet scattering theory\cite{Gre10}. We observe an excellent agreement for small transparencies, but eventually deviations appear for $D\gtrsim 0.3$. These discrepancies appear as the broadening of energy levels become so large that the effect of spurious emissions of (several) electron-hole pairs during a period becomes non-negligible. In this case, the mean number of emitted electrons during a period can exceed one. The counting statistics is important for characterizing the accuracy of the mesoscopic capacitor as a single electron emitter\cite{Albert2010}.

\section{Conclusions}
\label{sec:conc}

We have investigated experimentally and theoretically the
finite-frequency noise spectrum of the mesoscopic capacitor when
operated as a periodic single electron emitter. We have compared
experimental data with two complementary theoretical descriptions.
On one hand, we discussed the Floquet scattering theory which
allows us to accurately describe the system over the full range of
experimentally relevant parameters, in particular the energy
ranges in which charges are emitted. On the other hand, we
considered a semi-classical model which, despite its simplicity,
is able to account for the charge dynamics of the emitter when
operated under the optimal operating conditions. This model
allowed us to develop an analytic understanding of the measured
noise spectrum and the numerical results obtained using the
Floquet scattering theory.

Depending on the escape time of electrons from the mesoscopic
capacitor, two distinct noise regimes could be identified. When
the escape time is much smaller than the period of the drive, the
mesoscopic capacitor emits electrons and holes in an almost fully
periodic manner with unit probability and the main source of noise
is due to the uncertainty in the emission time within a period.
This type of noise, referred to as phase noise, can be clearly
identified both in the theoretical and experimental results. The
phase noise provides a fundamental lower limit on the noise,
arising from the random jitter between the triggering of emission
and the actual emission time. In the phase noise regime, we
obtained excellent agreement between our experimental data, the
Floquet scattering theory, and the semi-classical model. In the
other extreme, when the escape time of electrons is much larger
than the period of the drive, electron emission becomes rare and
the fluctuations are shot noise-like with a white spectrum
related to the average (particle) current $I_p = 2e f_d \times P$ through the usual
Schottky formula $S_{\mathrm{Schottky}} = 2eI_p$.

As the mesoscopic capacitor is tuned away from the optimal
operating conditions and charges are emitted close to the Fermi
energy, a significant increase in the noise is observed due to
additional charge fluctuations generated by the source. These
spurious emission processes are not accounted for by the
semi-classical model, in which maximally one electron--hole pair
can be emitted during each cycle. In contrast, these additional
fluctuations are fully accounted for by the Floquet scattering
theory. The ability to accurately investigate, model, and
characterize the single electron emission process will prove
useful in future few electron experiments involving the mesoscopic
capacitor as a controllable single electron source.

\acknowledgements

We thank P.~Degiovanni, Ch.~Grenier, G.\ Haack, and M.~Moskalets for instructive discussions. The work was supported by the Swiss NSF, the NCCR Quantum Science and Technology, the European Marie Curie Training Network NanoCTM, and the ANR grant ``1shot" (ANR-2010-BLANC-0412).

\end{document}